\documentclass[a4paper,11pt]{article}
\pdfoutput=1 

\usepackage{jheppub} 
\usepackage{array}
\usepackage{tikz}
\usepackage[normalem]{ulem}
\usetikzlibrary{decorations.pathmorphing}

\usepackage[T1]{fontenc} 

\newcommand{\bra}[1]{\langle #1 \rvert}
\newcommand{\ket}[1]{\lvert #1 \rangle}

\newcolumntype{C}[1]{>{\centering\arraybackslash}p{#1}}

\title{From CFT Data to Thermal Observables and Back Again in 3d Ising}

\author{A. Liam Fitzpatrick}
\author{and Noah Ring}
\affiliation{Department of Physics, Boston University,\\Boston, MA 02215, U.S.A.}

\emailAdd{fitzpatr@bu.edu}
\emailAdd{nmring@bu.edu}

\abstract{We investigate the relationship between thermal observables and new results for Conformal Field Theory (CFT) data in the 3d Ising CFT. On one hand, we explore the Eigenstate Thermalization Hypothesis (ETH) with subleading corrections and its relation to the low-dimension diagonal OPE coefficients of $\epsilon$, and compare to data for primaries with $\Delta < 15$ and $J\leq 4$.  We also compare the actual counts of primary states with various spins to predictions from a high-temperature analysis.  On the other hand, we explicitly perform Boltzmann sums over CFT data to construct thermal CFT observables on $S^2$, in particular the partition function and the thermal one-point function of $\epsilon$ on the sphere. By taking the high-temperature limit, we extract predictions for the thermal CFT parameters $f$ and $b_\epsilon$.
}

\begin{document} 
\maketitle
\flushbottom

\section{Introduction}\label{sec:intro}

    Conformal Field Theories (CFTs) are of broad importance in modern theoretical physics. An important feature of CFTs is that they are fully specified by their CFT data (i.e. the set of scaling dimensions and OPE coefficients), in that all observables are within reach if this data is known. Thus, in principle we may ``solve'' any particular CFT by collecting the full infinite set of its CFT data. Of course, truly collecting an infinite amount of data is intractable, especially since the high-energy spectrum of generic CFTs is expected to be chaotic. Instead, ``most'' CFT data involving high dimension operators will at best be accessible in terms of a statistical prediction.

    By studying these theories at finite temperature, both in flat space ($S^1_\beta\times\mathbb{R}^{d-1}$) and on the sphere ($S^1_\beta\times S^{d-1}$), we can leverage a number of new constraints to generate interesting relations between the CFT data and thermal observables. The Eigenstate Thermalization Hypothesis (ETH) in particular relates diagonal OPE coefficients of a primary operator to its thermal expectation value at an inverse temperature $\beta(\Delta)$ set by a thermodynamic function of the dimension of the external state, and predicts the statistical distribution of the off-diagonal OPE coefficients. Taking a scalar operator $\phi$ for simplicity, ETH predicts that at sufficiently high $\Delta$ (with all other quantum numbers such as spin held fixed), 
    \begin{align}
        f_{\phi\mathcal{OO}} \approx b_\phi\left(\frac{\Delta}{(d-1)\text{vol}(S^{d-1})f}\right)^{\Delta_\phi/d},
    \end{align}
    where $f_{\phi\mathcal{OO}}$ is a diagonal OPE coefficient, $\mathcal{O}$ is an operator with dimension $\Delta\gg 1$, $\Delta_\phi$ is the scaling dimension of $\phi$, $f$ parameterizes the flat space free energy density $F/V = -f\beta^{-d}$, and $b_\phi$ is a thermal vev parameterizing the flat space thermal expectation value of $\phi$
    \begin{align}
        \langle \phi \rangle_{S^1_\beta\times\mathbb{R}^{d-1}} &= \frac{b_\phi}{\beta^{\Delta_\phi}}.
    \end{align}
    The thermal data $f$ and $b_\phi$ can be computed from the zero-temperature CFT data. To see this relation, recall that the partition function and thermal one-point function on the sphere are computable as Boltzmann sums over the usual CFT data
    \begin{align}
        \mathcal{Z}(\beta) &= \sum_{\Delta}e^{-\beta\Delta}, & \langle\phi\rangle_{S^1_\beta\times S^{d-1}} &= \frac{1}{\mathcal{Z}(\beta)}\sum_{\Delta}\lambda_{\phi\mathcal{OO}}e^{-\beta\Delta}.
    \end{align}
    Then since in the infinite volume limit (equivalently the high-temperature limit) the sphere passes over to the plane, the thermal observables on the sphere must follow suit and match up with their flat space expectations.

    From the perspective of solving CFTs, it is interesting to understand not only how efficiently one can use the CFT data to predict thermal quantities, but also how efficiently one can go in the opposite direction and obtain approximations for some CFT data from thermal observables.  Perfect knowledge of one of these allows one to extract the other (e.g. the partition function contains the exact spectrum within it), but generally we will only have access to approximations of one or the other (for instance, a high-temperature expansion of the partition function).

    In order to explore the forward direction, from CFT data to thermal observables, we will directly construct thermal quantities as Boltzmann sums using available CFT data. By parameterizing the high-temperature regime using a thermal effective action \cite{Bhattacharyya:2007vs,Banerjee:2012iz,Jensen:2012jh,Benjamin:2023qsc,Benjamin:2024kdg,Delacretaz:2026owo,Kang:2022orq}, we can compare what we find with literature predictions. For the backwards direction, we will use ETH with subleading corrections \cite{Buric:2025uqt} and compare to existing data.
    
    In this work we will take the 3d Ising CFT as our example theory. The 3d Ising CFT is an ideal choice since it is a strongly coupled CFT which has a non-zero thermal mass and is believed to be ergodic, yet thanks to modern numeric techniques much of its CFT data is known. The recently developed Fuzzy Sphere regulator technology can be used to compute OPE coefficient data of the form $f_{\epsilon {\cal O}{\cal O}}$ at much higher dimensions $\Delta$ than was previously possible \cite{Zhu:2022gjc,Hu:2023xak,Fardelli:2024qla,Fardelli:2026zas}. In fact, for the purposes of this project, we used recent advancements in the Fuzzy Sphere \cite{Zhou:2026ugg} to compute 3d Ising CFT data for $\Delta < 15$ and $J\leq 4$. In total, for our study we obtained 319 primary operators, most of which were not previously known. Additionally, the 3d Ising CFT has already undergone previous finite temperature studies \cite{Iliesiu:2018fao,Iliesiu:2018zlz,Barrat:2025wbi,Barrat:2025nvu,PhysRevE.53.4414,PhysRevE.56.1642,PhysRevE.79.041142}, allowing us to compare our results with thermal data in the literature.
    
    The rest of the paper is organized as follows. In Section \ref{sec:Review} we briefly review relevant information about thermal one-point functions, the thermal effective action, and ETH. In Section \ref{sec:FSandETH} we compare the spectrum of diagonal OPE coefficients in the 3d Ising CFT to the na\"ive predictions of ETH at leading order in the large-$\Delta$ limit. We then take into account a leading universal correction to ETH and check how well it describes the diagonal OPE coefficients. 
    
    In Section \ref{sec:FiniteTObservablesOnSphere} we review how the 3d Ising CFT partition function on the sphere is constructed from the CFT data by a direct Boltzmann sum, and how one can extract a value for the scheme-independent part of the free energy density $f$. We then use the CFT data from the Fuzzy Sphere to do a similar construction of the thermal one-point function of $\epsilon$ on the sphere by a direct Boltzmann sum (using the thermal one-point conformal blocks to resum conformal descendants) and extract a value for the thermal vev $b_\epsilon$. 
    
    In Section \ref{sec:ETHCorrections} we return to ETH in order to explore universal subleading corrections. As a warm-up we consider the asymptotic density of primary states and compute a tower of subleading universal terms. We show that we can resum all of these universal terms and arrive at a closed-form expression for the density of primary states. We repeat this exercise for ETH and compute a tower of subleading universal corrections to compare with the diagonal OPE coefficients. We then Borel resum the asymptotic expansion and compare with the diagonal OPE data once more. We conclude in Section \ref{sec:SummaryAndFutureDirec} with a summary of our results and discussion of future directions.

\section{Finite Temperature CFT and ETH Review}\label{sec:Review}

    We will study finite (non-zero) temperature CFTs in both flat space ($S^1_\beta\times\mathbb{R}^{d-1}$) and on the sphere ($S^1_\beta\times S^{d-1}$) in $d=3$ dimensions. For the following review however we will keep $d$ general, but assume $d>2$. We are specifically interested in thermal one-point functions in this work, so we will make no comments on higher-point correlators. For recent work on thermal two-point functions in both general and holographic CFT settings, see \cite{Iliesiu:2018fao,Iliesiu:2018zlz,Barrat:2025nvu,Buric:2025anb,Marchetto:2023xap,Barrat:2025wbi,Alday:2020eua,Dodelson:2022yvn,Karlsson:2022osn,Dodelson:2023vrw,Dodelson:2020lal,Dodelson:2023nnr,Barrat:2025twb,Buric:2025fye,Barrat:2024aoa}. Finally, we will use coordinates $x = (\tau,\vec{x})$ for flat space and $x = (\tau,\hat{n})$ for the sphere, with the identification $\tau \sim \tau + \beta$ for the thermal circle.

    \subsection{Thermal One-Point Functions in CFTs}\label{sec:FiniteTCFTs}

    In a finite temperature CFT in flat space,  only even-spin primary operators in symmetric traceless representations have non-zero thermal one-point functions.\footnote{Additional global symmetries of the CFT can also force certain one-point functions to vanish. For example, in the 3d Ising CFT the primary operator $\sigma$ is odd under the global $\mathbb{Z}_2$ symmetry which forces its thermal one-point function to vanish identically.} These flat space thermal one-point functions are constrained to take the following form
    \begin{align}
        \langle\mathcal{O}^{\mu_1\mu_2\cdots\mu_\ell}(x)\rangle_{S^1_\beta\times\mathbb{R}^{d-1}} &= \frac{b_\mathcal{O}}{\beta^{\Delta_\mathcal{O}}}(e^{\mu_1}e^{\mu_2}\cdots e^{\mu_\ell} - \text{traces}), \label{eq:flatspace1ptfunc}
    \end{align}
    where $b_\mathcal{O}$ is an undetermined operator and theory dependent coefficient and $e^\mu$ is a unit vector pointed in the Euclidean time direction. We will sometimes refer to $b_\mathcal{O}$ as a thermal vev.

    If we instead consider a finite temperature CFT on the sphere, conformal invariance is less constraining on the form of the one-point function. For a sphere of radius $R$, the one-point function can at best be constrained to take the form\footnote{Other references write $h\left(\frac{\beta}{R}\right) = b_\mathcal{O}f\left(\frac{\beta}{R}\right)$ owing to the sphere to flat space limit. We write the one-point function in this alternative way for later convenience, but both conventions are equivalent.}
    \begin{align}
        \langle\mathcal{O}^{\mu_1\mu_2\cdots\mu_\ell}(x)\rangle_{S^1_\beta\times S^{d-1}_R} &= \frac{h\Big(\frac{\beta}{R}\Big)}{\beta^{\Delta_\mathcal{O}}}(e^{\mu_1}e^{\mu_2}\cdots e^{\mu_\ell} - \text{traces}), \label{eq:sphere1ptfunc}
    \end{align}
    where $h\left(\frac{\beta}{R}\right)$ is a function of the dimensionless ratio $\frac{\beta}{R}$ that depends on ${\cal O}$. By inspection, the high-temperature limit $\beta \rightarrow 0$ and the flat space limit $R \rightarrow \infty$ are equivalent:
    \begin{align}
        \lim_{R\to\infty}  \beta^{\Delta_{\cal O}} \langle\mathcal{O}^{\mu_1\mu_2\cdots\mu_\ell}(x)\rangle_{S^1_\beta\times S^{d-1}_R} = \lim_{\beta\to 0}  \beta^{\Delta_{\cal O}} \langle\mathcal{O}^{\mu_1\mu_2\cdots\mu_\ell}(x)\rangle_{S^1_\beta\times S^{d-1}_R}.
    \end{align}
    From here on, we return to units where $R=1$, and let $\beta\to 0$ act as the relevant flat space limit. Since we must recover the flat space one-point function in the high-temperature limit,  the unknown function $h(\beta)$ necessarily satisfies the boundary condition $h(0) = b_\mathcal{O}$.

    Besides being naturally related to the flat space one-point function by a high-temperature limit, the one-point function on the sphere admits a simple thermal trace definition owing to the state-operator correspondence. Taking a scalar primary operator $\phi$ for notational simplicity, we can write the thermal one-point function on the sphere as
    \begin{align}
        \langle \phi(x)\rangle_{S^1_\beta\times S^{d-1}} &= \frac{1}{\mathcal{Z}(\beta)}\text{tr}\left(\phi(x) e^{-\beta D}\right) = \frac{1}{\mathcal{Z}(\beta)}\sum_{\mathcal{O}}\bra{\mathcal{O}}\phi(x)\ket{\mathcal{O}}e^{-\beta\Delta_{\mathcal{O}}},
    \end{align}
    where $\mathcal{O}$ is any local operator in the CFT and $\mathcal{Z}(\beta)$ is the CFT partition function on the sphere
    \begin{align}
        \mathcal{Z}(\beta) &= \text{tr}\left(e^{-\beta D}\right) = \sum_{\mathcal{O}} e^{-\beta\Delta_\mathcal{O}}.
    \end{align}
    Importantly, the state-operator correspondence tells us that the sum over expectation values can be understood as a sum over OPE coefficients
    \begin{align}
        \langle \phi(x)\rangle_{S^1_\beta\times S^{d-1}} &= \frac{1}{\mathcal{Z}(\beta)}\sum_{\mathcal{O}}\lambda_{\phi\mathcal{OO}}e^{-\beta\Delta_{\mathcal{O}}},
    \end{align}
    implying that the thermal one-point function on the sphere is entirely determined by zero-temperature CFT data. The high-temperature limit then relates this  calculation to the flat space thermal one-point function
    \begin{align}
        \frac{1}{\mathcal{Z}(\beta)}\sum_{\mathcal{O}}\lambda_{\phi\mathcal{OO}}e^{-\beta\Delta_{\mathcal{O}}} \xrightarrow{\beta\to 0}\frac{b_\phi}{\beta^{\Delta_\phi}}.
    \end{align}
    Therefore, we have a formal procedure for computing $b_\phi$ as the following limit
    \begin{align}
        \lim_{\beta\to 0}\beta^{\Delta_\phi}\langle \phi(x)\rangle_{S^1_\beta\times S^{d-1}} &=  \lim_{\beta\to 0}\frac{\beta^{\Delta_\phi}}{\mathcal{Z}(\beta)}\sum_{\mathcal{O}}\lambda_{\phi\mathcal{OO}}e^{-\beta\Delta_{\mathcal{O}}} = b_\phi.
    \end{align}
    Finally, thermal vevs appear as inputs when evaluating thermal two-point functions using the OPE. To see this, let's take the following thermal two-point function in flat space as an example
    \begin{align}
        \langle \phi(x) \phi(0)\rangle_{S^1_\beta\times\mathbb{R}^{d-1}}.
    \end{align}
    As long as $\lvert x \rvert = \sqrt{(\tau^2 + \vec{x}^{\,2})} < \beta$, then we can expand it using the OPE
    \begin{align}
        \langle \phi(x) \phi(0)\rangle_{S^1_\beta\times\mathbb{R}^{d-1}} &= \sum_{\mathcal{O}\in\phi\times\phi} \frac{f_{\phi\phi\mathcal{O}}}{c_\mathcal{O}}\frac{b_\mathcal{O}}{\beta^{\Delta_\mathcal{O}}}\frac{J!}{2^J(\nu)_J}C_J^{(\nu)}\left(\frac{\tau}{\lvert x \rvert}\right)\lvert x \rvert^{\Delta_\mathcal{O}-2\Delta_\phi},
    \end{align}
    where $c_\mathcal{O}$ is the normalization for the zero-temperature two-point function of $\mathcal{O}$, $f_{\phi\phi\mathcal{O}}$ is an OPE coefficient, $\nu = \frac{d-2}{2}$, $(a)_n$ is the Pochhammer symbol, and $C^{(\alpha)}_n(x)$ is a Gegenbauer polynomial.

    \subsection{The Thermal Effective Action}\label{sec:ThermalEFT}

    In order to study finite temperature CFTs we compactify the Euclidean time direction $\mathbb{R}$ to the thermal circle $S^1$ with circumference $\beta$.\footnote{Note that for this section only we leave $R\neq 1$ as well as rescale our coordinates such that $\tau \sim \tau + 1$ in order to make contact with the existing literature.} Typically this compactification will lead to a gapped $(d-1)$-dimensional theory with thermal mass $m_\text{th}\sim 1/\beta$.\footnote{Sometimes this compactification will not lead to a fully gapped theory, e.g.~if there are modes which are protected by symmetry.} The existence of this gap means that equilibrium thermal correlators will exponentially decay at large spatial distances
    \begin{align}
        \langle\mathcal{O}(0,\vec{x})\mathcal{O}(0)\rangle_\beta \sim e^{-m_\text{th}\lvert\vec{x}\rvert}.
    \end{align}
    This decay implies that if we couple our theory to classical background fields and integrate out the gapped modes, we will be left with a local effective action in the background fields which we refer to as a thermal effective action.

    Of interest for us here will be understanding how finite temperature CFT observables on the sphere $S^{d-1}_R$ are related to their flat space $\mathbb{R}^{d-1}$ analogs in the high-temperature limit. We can understand the local response by coupling to a $d$-dimensional background metric 
    \begin{align}
        ds^2 &= g_{\mu\nu}(x)dx^\mu dx^\nu = \tilde{g}_{ij}(x)dx^idx^j + e^{2\sigma(x)}(d\tau + A_i(x)dx^i)^2,\label{eq:KKMetric}
    \end{align}
    where the above metric has been put into Kaluza-Klein (KK) form with $x^0=\tau$ being the periodic direction such that $\tau\in[0,1)$. The fields in the $(d-1)$-dimensional space are a metric $\tilde{g}_{ij}$, the KK gauge field $A_i$, and a dilaton field $\sigma$.\footnote{Note that when we restrict to the sphere, we will be able to identify $e^\sigma = \beta = \text{constant}$.} Thus, the $d$-dimensional CFT defined on (\ref{eq:KKMetric}) has a partition function given by
    \begin{align}
        \mathcal{Z}_\text{CFT}[g] = \int D\psi\,e^{-S[\psi,\tilde{g},A,\sigma]} \equiv e^{-S_\text{th}[\tilde{g},A,\sigma]},
    \end{align}
    where $\psi$ is a stand-in for the dynamical fields of the CFT that we have integrated out to define the local generating functional $S_\text{th}[\tilde{g},A,\sigma]$, i.e. the thermal effective action.

    We can now constrain what operators can appear in the thermal effective action with symmetry. The $d$-dimensional coordinate invariance requires that the thermal effective action is invariant under $(d-1)$-dimensional coordinate transformations and gauge transformations of the KK gauge field. Weyl invariance inherited from the CFT then imposes further constraints on our thermal effective action. The $\tau$-independent Weyl transformations act on the metric (\ref{eq:KKMetric}) as
    \begin{align}
        \tilde{g}_{ij} &\to e^{2\phi}\tilde{g}_{ij}, & A_i &\to A_i, & \sigma &\to \sigma + \phi.
    \end{align}
    We can therefore eliminate the $\sigma$ dependence in $S_\text{th}$ by setting $\phi = -\sigma$, giving us 
    \begin{align}
        S_\text{th}[\tilde{g},A,\sigma] &= S_\text{th}[e^{-2\sigma}\tilde{g},A,0] - S_\text{anom}[\tilde{g},A,\sigma;-\sigma],
    \end{align}
    where $S_\text{anom}$ is the contribution from the Weyl anomaly. Our primary focus in this work will be $d=3$ dimensional CFTs where this anomaly is absent. See \cite{Benjamin:2023qsc} for a more general discussion of Weyl anomaly terms in CFT thermal effective actions. If we now define a rescaled spatial metric 
    \begin{align}
        \hat{g}_{ij} &\equiv e^{-2\sigma}\tilde{g}_{ij},
    \end{align}
    we can write down a thermal effective action in terms of diff-invariant and gauge invariant objects using $\hat{g}_{ij}$ and $A_i$
    \begin{align}
        S_\text{th}[\hat{g},A] &= \int d^{d-1}x\sqrt{\hat{g}}\left(-f + c_1\hat{\mathcal{R}} + c_2 F^2 + \cdots \right),\label{eq:GenThermEffAction}
    \end{align}
    where $\hat{\mathcal{R}}_{ijkl}$ is the Riemann tensor built out of the rescaled metric $\hat{g}_{ij}$, $F_{ij}$ is the field strength for the KK gauge field, and all contractions are done with $\hat{g}_{ij}$. This thermal EFT is a derivative expansion where each derivative is implicitly suppressed by $1/\Lambda_\text{UV} = 1/m_\text{th} \sim \beta$. If we further restrict to parity preserving CFTs, then by dimensional analysis, all allowed operators have even dimensions. Thus, the action (\ref{eq:GenThermEffAction}) computes corrections to observables in an expansion of even powers of $\beta$ relative to a leading power. Restricting to the case where our spatial manifold is $S^{d-1}_R$, the thermal effective action behaves as 
    \begin{align}
        S_\text{th} \sim \text{vol}(S^{d-1})R^{d-1}\beta^{-(d-1)}\left[-f + \#\left(\frac{\beta}{R}\right)^{2} + \# \left(\frac{\beta}{R}\right)^4 + \cdots\right].
    \end{align}
    Therefore, these corrections translate into a high-temperature series expansion for the free energy density
    \begin{align}
        \frac{F}{\text{vol}(S^{d-1}_R)} &= \beta^{-d}\left[-f + \#\left(\frac{\beta}{R}\right)^{2} + \# \left(\frac{\beta}{R}\right)^4 + \cdots\right],
    \end{align}
    where $F=-\beta^{-1}\log\mathcal{Z}$ as usual and we can identify $f$ as the scheme-independent part of the free energy density.

    We can extend this construction to tell us about the geometric response induced in thermal one-point functions. For simplicity let us consider a scalar primary operator $\mathcal{O}$. Further, assume that $\mathcal{O}$ carries no non-trivial quantum numbers under any symmetry of the theory that is preserved by the thermal state. This restriction implies that $\mathcal{O}$ can have a non-zero flat space thermal one-point function, which is important since we want to understand how the sphere one-point function approaches the flat space analog in the $\beta\to 0$ limit. We can source this operator with a background scalar field $\lambda(x)$ by taking
    \begin{align}
        S \to S + \int d^dx\, \sqrt{g} \lambda(x)\mathcal{O}(x),
    \end{align}
    which means that now
    \begin{align}
        \mathcal{Z}_\text{CFT}[g,\lambda] = \int D\psi\,e^{-S[\psi,\tilde{g},A,\sigma,\lambda]} \equiv e^{-S_\text{th}[\tilde{g},A,\sigma,\lambda]}.
    \end{align}
    The same symmetry arguments from before continue to hold, with $\lambda$ transforming under $\tau$-independent Weyl transformations as $\lambda \to e^{(\Delta-d)\phi}\lambda$. So for our choice of $\phi = -\sigma$, we can define a rescaled background field
    \begin{align}
        \hat{\lambda} \equiv e^{-(\Delta-d)\sigma}\lambda.
    \end{align}
    To linear order in  $\hat{\lambda}$, the new terms in the thermal effective action can be parameterized as 
    \begin{align}
        S_\text{th}[\hat{g},A,\hat{\lambda}] \supset \int d^{d-1}x\sqrt{\hat{g}}\hat{\lambda} \left(b_\mathcal{O} + b_1\hat{\mathcal{R}} + b_2F^2 + \cdots\right).
    \end{align}
    Restricting again to $S^{d-1}_{R}$, the same dimensional analysis arguments as before show us that the thermal effective action predicts that the thermal one-point function on the sphere in the high-temperature limit will behave as
    \begin{align}
        \langle\mathcal{O}\rangle_{S^1\times S^{d-1}_R} &= -\frac{\beta^{d-\Delta}}{\text{vol}(S^1\times S^{d-1}_R)}\frac{\partial \log\mathcal{Z}}{\partial \hat{\lambda}} = \frac{1}{\beta^{\Delta}}\left[b_\mathcal{O} + \# \left(\frac{\beta}{R}\right)^2  + \#\left(\frac{\beta}{R}\right)^4 + \cdots\right].
    \end{align}

    We now would like to restrict to $d=3$ and briefly discuss the effect of turning on a non-zero chemical potential $\mu$ for angular momentum $J$. Turning on $\mu$ leads to twisted boundary conditions on $S^1 \times S^2_R$ such that $(\tau,\varphi)\sim(\tau + 1,\varphi + \mu)$. If we work with co-rotating coordinates $\varphi' = \varphi - \mu\tau$, chosen so that $\varphi'$ is identified with itself after going around the thermal circle, it is easy to read off from the $d\tau^2$ term in the new form of the metric that
    \begin{align}
        e^{2\sigma} &= \beta^2 + \mu^2 R^2\sin^2\theta.
    \end{align}
    It  follows that at leading order, the partition function takes the form
    \begin{align}
        \mathcal{Z} &= e^{\frac{4\pi R^2 f}{\beta^2 + \mu^2R^2}}.
    \end{align}
    Similarly, at leading order the thermal one-point function takes the form
    \begin{align}
        \langle \mathcal{O}(x)\rangle_{S^1\times S^2_R} &= \frac{b_\mathcal{O}}{(\beta^2+\mu^2 R^2\sin^2\theta)^{\Delta/2}}.
    \end{align}

    In this work we will not consider non-perturbative corrections to the thermal effective action; for a recent discussion, see \cite{Benjamin:2023qsc}.

    \subsection{The Eigenstate Thermalization Hypothesis}

    In a quantum field theory (QFT), the Eigenstate Thermalization Hypothesis (ETH) tells us that thermal correlation functions of local operators are well-approximated by zero-temperature correlation functions of those same local operators in the background of highly excited states
    \begin{align}
        \langle \mathcal{O}_1\mathcal{O}_2\cdots\mathcal{O}_n \rangle_\beta \approx \bra{E} \mathcal{O}_1\mathcal{O}_2\cdots\mathcal{O}_n\ket{E}.
    \end{align}
    More precisely, when restricting to matrix elements of an individual operator, ETH says that \cite{Srednicki_1999,Fardelli:2026zas}
    \begin{align}
        \bra{E_\alpha}\mathcal{O}\ket{E_\beta} &= \langle \mathcal{O} \rangle_{\beta(E)}\delta_{\alpha\beta} + e^{-S(E)/2}f_{\mathcal{O}}(E,\omega)R_{\alpha\beta},
    \end{align}
    where $\alpha$ and $\beta$ label possibly distinct energy eigenstates, $E \equiv \frac{1}{2}(E_\alpha+E_\beta)$ and $\omega \equiv E_\alpha-E_\beta$, $\langle \mathcal{O} \rangle_{\beta(E)}$ is the  thermal expectation value of $\mathcal{O}$, $\beta(E)$ is the inverse temperature at energy $E$, $S(E)$ is the thermodynamic entropy at energy $E$, $f_{\mathcal{O}}(E,\omega)$ is a smooth function fixed by the thermal two-point function of $\mathcal{O}$, and $R_{\alpha\beta}$ is an element of a random matrix with zero mean and unit variance.

    This relation has interesting consequences in the context of CFTs. There, diagonal OPE coefficients  of primary operators are  translated by ETH into a statement about thermal expectation values. For instance, for a scalar operator $\phi$, ETH says that 
    \begin{align}
        f_{\phi\mathcal{OO}} \equiv \frac{1}{\text{vol}(S^{d-1})}\int_{S^{d-1}}\bra{\mathcal{O}}\phi(\hat{n})\ket{\mathcal{O}} = \langle \phi \rangle_{S^1_\beta\times S^{d-1}}\Big\vert_{\beta=\beta(\Delta)} + \mathcal{O}(e^{-S(\Delta)/2}), \label{eq:ETHchain}
    \end{align}
    where we are averaging $\phi(\hat{n})$ over $S^{d-1}$ and $\Delta$ is the dimension of $\mathcal{O}$; when the operator ${\cal O}$ has spin, and therefore the primary `state' is a full multiplet of spin, the integral above projects onto the diagonal component of the OPE coefficients.
    
    In order to fully fix the ETH relation, we require the thermodynamic function $\beta(\Delta)$. We recall that the free energy of a $d$-dimensional CFT is fixed to be 
    \begin{align}
        F &= -V f\beta^{-d},
    \end{align}
    where $V$ is the volume of space and $f$ is a theory-dependent constant, sometimes referred to as the scheme-independent part of the free energy density. Using the usual thermodynamic relations
    \begin{align}
        S &= \beta^2\partial_\beta F, & E &= \partial_\beta(\beta F),
    \end{align}
    it follows that $\beta(\Delta)$ is given by
    \begin{align}
        \beta &= \left(\frac{(d-1)\text{vol}(S^{d-1})f}{\Delta}\right)^{1/d}, \label{eq:dDimTempEnergy}
    \end{align}
    where we have taken $V = \text{vol}(S^{d-1}) = \frac{2\pi^{d/2}}{\Gamma(\frac{d}{2})}$. This relation is only valid in the thermodynamic limit $\Delta\gg1$. Taking $\mathcal{O}$ to be such a heavy operator and plugging the flat space thermal one-point function (\ref{eq:flatspace1ptfunc}) into (\ref{eq:ETHchain}) and evaluating at the inverse temperature (\ref{eq:dDimTempEnergy}),  ETH predicts that OPE coefficients in a $d$-dimensional CFT should asymptotically behave as 
    \begin{align}
        f_{\phi\mathcal{OO}} &= b_\phi\left(\frac{\Delta}{(d-1)\text{vol}(S^{d-1})f}\right)^{\Delta_\phi/d} \ \stackrel{d=3}{=} \  b_\phi\left(\frac{\Delta}{8\pi f}\right)^{\Delta_\phi/3}.
         \label{eq:3dETH}
    \end{align}

\section{Fuzzy Sphere Data and Leading Order ETH}\label{sec:FSandETH}

    The recent development of the Fuzzy Sphere \cite{Zhu:2022gjc} as a non-perturbative tool for studying 3d CFTs has opened up a new direction for the extraction of data from all kinds of CFTs. An important advantage that the Fuzzy Sphere provides is direct study of a CFT of interest on $S^2$, which by the state-operator correspondence means that all of the CFT Hamiltonian eigenstates are in one-to-one correspondence with CFT operators. OPE coefficients then become easily accessible since they are literally matrix elements of local operators evaluated in eigenstates of the Hamiltonian. See \cite{He:2026ong} for a comprehensive review with references therein.

    Recently, this method was used to calculate many new OPE coefficients of the form $\langle {\cal O}_i | \epsilon | {\cal O}_j\rangle$ in the 3d Ising CFT, up to operators ${\cal O}_i$ with dimension $\Delta\lesssim 9$ \cite{Fardelli:2026zas}. The diagonal elements of $\epsilon$ are of particular interest because ETH specifically predicts that
    \begin{align}
        f_{\epsilon\mathcal{OO}} &= \frac{1}{\sqrt{4\pi}}\int d\Omega\,Y_{0,0}\bra{\mathcal{O}}\epsilon(\hat{n})\ket{\mathcal{O}} = b_\epsilon \left(\frac{\Delta}{8\pi f}\right)^{\Delta_\epsilon/3},\label{eq:ETHPRediction}
    \end{align}
    where $\Delta \gg 1$ is the scaling dimension of $\mathcal{O}$ and $\Delta_\epsilon = 1.412625(10)$ \cite{Kos:2016ysd} is the scaling dimension of $\epsilon$. Having this collection of low-dimension OPE coefficient data allows us to directly compare to the leading ETH prediction and see how well it describes this part of the spectrum. 
     
    Ideally, one would like data for even heavier operators than $\Delta \lesssim 9$. The primary limitation in \cite{Fardelli:2026zas} was the rapid growth of the CFT Hilbert space, making further spectral depth too expensive. This issue was recently alleviated by the ability to construct an $SO(3)$-resolved basis in the canonical Fuzzy Sphere code FuzzifiED \cite{Zhou:2026ugg}. The authors probed the 3d Ising spectrum in the $J\leq 2$ sectors, reaching dimensions as high as $\Delta \lesssim 16.5$ in some cases. Therefore, in order to make all comparisons more robust, we have computed the dimensions $\Delta$ and diagonal OPE coefficients $f_{\epsilon\mathcal{OO}}$ for primary operators with $\Delta < 15$ and $J \leq 4$ using the methods of \cite{Zhou:2026ugg}. Details are provided in Appendix \ref{app:FuzzySphereData}.

\begin{figure}[h!]
        \centering
        \includegraphics[width=0.9\linewidth]{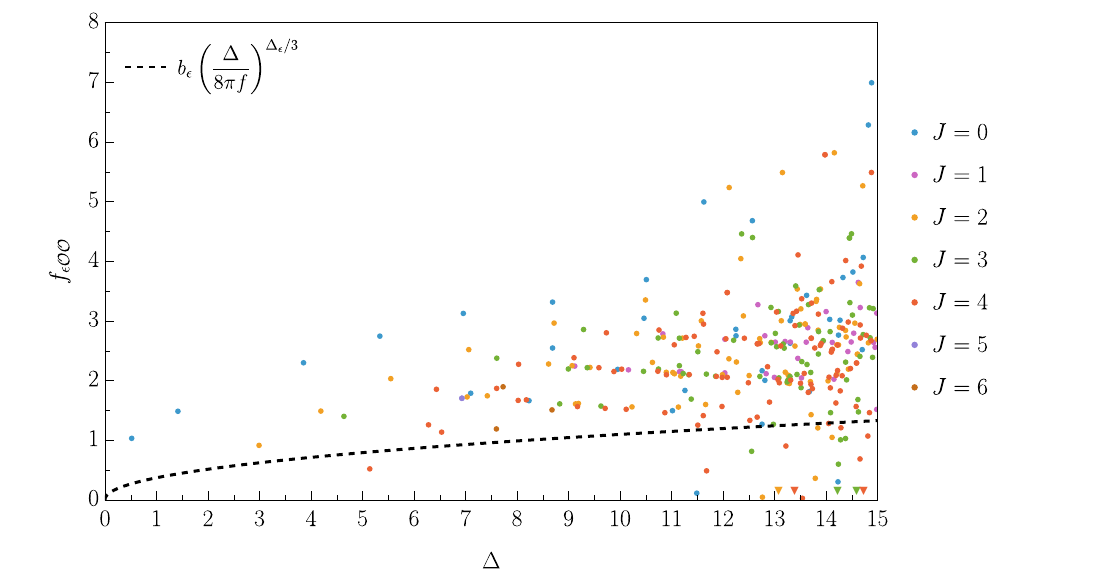}
        \caption{ETH plotted as a function of $\Delta$ against the diagonal OPE coefficients of $\epsilon$. Each point corresponds to an OPE coefficient with a different primary operator $\mathcal{O}$ and the points are color coded by the spin of $\mathcal{O}$. Arrows indicate a point outside of frame. The ETH prediction is represented by the black dashed line. Error bars are suppressed for readability.}
        \label{fig:LeadingNaiveETH}
    \end{figure}

    With this larger set of diagonal OPE data, we now want to compare with ETH. Taking $b_\epsilon$ and $f$ to be the following values from previous estimates in the literature \cite{Iliesiu:2018zlz,Barrat:2025wbi,PhysRevE.79.041142,PhysRevE.53.4414,PhysRevE.56.1642,Benjamin:2023qsc}
    \begin{align}
        b_\epsilon &\approx 0.7, & f &\approx 0.153,
    \end{align}
    we plot (\ref{eq:ETHPRediction}) against the diagonal OPE coefficients of $\epsilon$ in Fig.~\ref{fig:LeadingNaiveETH}, as was done in \cite{Fardelli:2026zas} but now over a larger range in $\Delta$.\footnote{Note that error bars have been omitted from Fig.~\ref{fig:LeadingNaiveETH} for readability, and arrows indicate that there is an OPE coefficient which lies outside of the frame.} We observe in Fig.~\ref{fig:LeadingNaiveETH} that there is clear disagreement between ETH and the available diagonal OPE coefficients of $\epsilon$ up to $\Delta < 15$. In particular, most of the OPE coefficients lie above the ETH prediction. ETH is expected to hold in the thermodynamic/asymptotic regime of the theory when the external dimensions $\Delta$ are large, so this disagreement is not a contradiction per se, but one of our primary goals will be to understand the disagreement better and see if we can improve on it.

    As a first step toward this goal, we can look at similar work which was done with the free scalar theory \cite{Buric:2025uqt}. There, the authors looked at the operator $\phi^2$ and its diagonal OPE coefficients $f_{\phi^2\mathcal{OO}}$\footnote{In \cite{Buric:2025uqt} the authors choose a different tensor structure basis which implies that their OPE coefficients are not exactly the same as the harmonic projected OPE coefficients in \cite{Fardelli:2026zas}. However, what the authors of \cite{Buric:2025uqt} refer to as the `dominant' OPE coefficient $\overline{\lambda^{a=0}_{\phi\mathcal{OO}}}$ turns out to be exactly equal to the sphere averaged OPE coefficient relevant for ETH: $f_{\phi\mathcal{OO}} = \overline{\lambda^{a=0}_{\phi\mathcal{OO}}}$.} and compared them to ETH as well as ETH plus subleading corrections (see \cite{Buric:2025uqt} Fig. 6 and Fig. 7). In their plots, the authors found that the $f_{\phi^2\mathcal{OO}}$ OPE coefficients converged to the ETH plus subleading corrections curve at different $\Delta$ depending on the spin of the external state $\mathcal{O}$. For example, $f_{\phi^2\mathcal{OO}}$ for spinning external states converged to ETH plus three subleading orders around $\Delta\approx 15$, while the $f_{\phi^2\mathcal{OO}}$ for scalar external states took until about $\Delta \approx 20$ to converge to the same curve. Although the finite temperature free scalar theory has some quantitative differences, it is possible that this behavior is qualitatively general and we can expect something similar to occur in the 3d Ising CFT.

    Motivated by these observations, we can take into account just the first subleading correction to ETH which we compute in Section \ref{sec:UniversalSubleadingETH} using the machinery of \cite{Buric:2025uqt}
    \begin{align}
        f_{\phi\mathcal{OO}} &= b_\phi\left(\frac{\Delta}{8\pi f}\right)^{\Delta_\phi/3}\left[1 + \frac{\Delta_\phi(6-\Delta_\phi)}{3\pi^{1/3} f^{1/3}\Delta^{2/3}} + \mathcal{O}(\Delta^{-1})\right].\label{eq:ETHwithSubleading}
    \end{align}
    Note that this first subleading term differs from the one computed in \cite{Buric:2025uqt}, with the difference stemming from an alternative choice of the form of the scalar thermal one-point function at non-zero chemical potential.

    \begin{figure}[h!]
        \centering
        \includegraphics[width=0.9\linewidth]{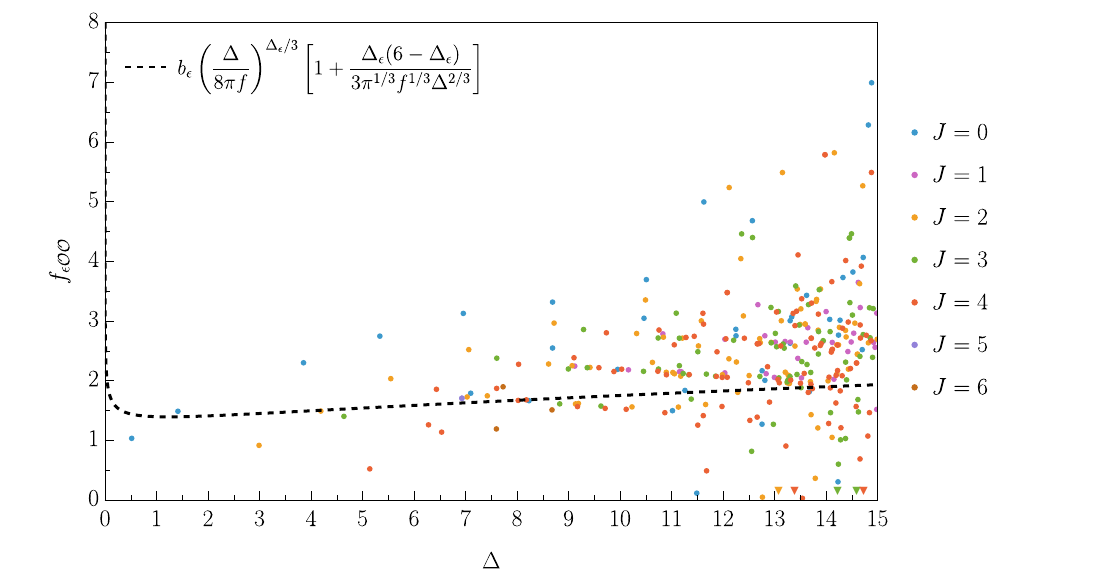}
        \caption{ETH plus a subleading correction plotted as a function of $\Delta$ against the diagonal OPE coefficients of $\epsilon$. Each point corresponds to an OPE coefficient with a different primary operator $\mathcal{O}$ and the points are color coded by the spin of $\mathcal{O}$. Arrows indicate a point outside of frame. The ETH plus a subleading correction curve is represented by the black dashed line. Error bars are suppressed for readability.}
        \label{fig:SubleadingNaiveETH}
    \end{figure}
    
    Taking $\phi = \epsilon$, we plot (\ref{eq:ETHwithSubleading}) with the $f_{\epsilon\mathcal{OO}}$ OPE data in Fig. \ref{fig:SubleadingNaiveETH}. We again find disagreement between the data and the asymptotic prediction. Although the curve is now closer to the value of many more of the OPE coefficients, it does not obviously predict or describe any of them. Indeed, the coefficient of the subleading term is actually $\mathcal{O}(1)$, even if we evaluate $\Delta_\phi$ at the unitarity bound. For the $\Delta_\phi = \Delta_\epsilon$ case that we are interested in, this subleading contribution is the dominant term until about $\Delta\approx 5$. This fact suggests that higher-order terms are likely necessary before there is any noticeable agreement for $\Delta < 15$.\footnote{Computing microcanonical averages of the data also does not improve agreement with the leading and subleading ETH computations. However, such averaging will be important when we go to much higher order in Section \ref{sec:UniversalSubleadingETH}.}

    Thus, the observations we have made so far make it clear that the available diagonal OPE data is not yet in a regime which is described by leading-order ETH or ETH with a single subleading term. Given these results, a more relevant question to ask is whether or not the available data can appropriately describe any of the thermal physics. Thus, we would like to switch gears and directly compute thermal CFT quantities on $S^2$ using our limited set of CFT data, with the goal of checking how well thermal quantities computed from the available CFT data adhere to large volume thermodynamic expectations.

\section{Thermal Observables on \texorpdfstring{$S^2$}{S2}}\label{sec:FiniteTObservablesOnSphere}

    We will now explicitly compute thermal CFT observables on $S^1_\beta\times S^2$. The two sphere observables we will compute are the partition function and the thermal one-point function of $\epsilon$: $\mathcal{Z}(\beta)$ and $\langle\epsilon\rangle_{S^1_\beta\times S^2}$. To our knowledge, this is the first computation of  the expectation value $\langle\epsilon\rangle_{S^1_\beta\times S^2}$ at finite temperature on the sphere as a function of $\beta/R$. Our goal is to extract the parameters $f$ (the scheme-independent part of the free energy density) and $b_\epsilon$ (the thermal vev of $\epsilon$) which parameterize these objects in the $\beta\to 0$ limit. The idea is to rely on the thermal effective action to predict the small-$\beta$ behavior so that we can extract $f$ and $b_\epsilon$ respectively. By comparing what we find with the literature, we will be able to test how well we are accessing the high-temperature regime.

    Although in principle an infinite amount of CFT data must be summed over in order to exactly carry out these limits, we find that a finite amount of data is sufficient for accessing a regime of $\beta$ where the thermal effective action applies. In our calculations we will use the complete spectrum below $\Delta \lesssim 9$ and the spectrum below $\Delta < 15$ with $J\leq 4$. We refer to this as the ``combined'' CFT data. We compare results from using the combined spectrum to results from using just the spectrum below $\Delta\lesssim 9$ (i.e. the spectrum known to \cite{Fardelli:2026zas}) in order to understand how advantageous our expanded but incomplete spectrum is.

    \subsection{The Partition Function and \texorpdfstring{$f$}{f}}\label{sec:PartFunc}

    Our goal is to approximately construct the 3d Ising CFT partition function on the sphere and check if we can probe into the high-temperature regime. Recall that any CFT partition function on the sphere can be computed as a simple sum over Boltzmann weights 
    \begin{align}
        \mathcal{Z}(\beta) &= \sum_{\Delta} e^{-\beta\Delta},\label{eq:PartitionFuncDefn}
    \end{align}
    where $\Delta$ goes over all states in the CFT, both primaries and descendants. However, since we will be working with a limited amount of data, we will be able to go much further if we instead decompose the partition function into a sum over characters
    \begin{align}
        \mathcal{Z}(\beta) &= \sum_{\Delta\text{ primary}}\chi_{\Delta,J}(\beta),\label{eq:PartitionFuncChiDefn}
    \end{align}
    where $\chi_{\Delta,J}(\beta)$ is the 3d character for a primary operator of dimension $\Delta$ and spin $J$\footnote{Note that any conserved currents such as the stress tensor have their shortened multiplets accounted for with the short character $\chi^\text{short}_J(\beta) = \chi_{J+1,J}(\beta) - \chi_{J+2,J-1}(\beta)$.}
    \begin{align}
        \chi_{\Delta,J}(\beta) &= \frac{(2J+1)e^{-\beta\Delta}}{(1-e^{-\beta})^3}.
    \end{align}
    Thus, for every primary operator that we add to our approximate partition function, the infinite tower of its descendant contributions will be accounted for. The thermal effective action from Section \ref{sec:ThermalEFT} tells us that in the high-temperature limit the CFT partition function on the sphere should behave as
    \begin{align}
        \mathcal{Z}_{\text{high-}T}(\beta) &\approx e^{\frac{4\pi f}{\beta^2}},
    \end{align}
    where $f$ is the scheme-independent part of the free energy density. Importantly, the parameter $f$ controls much of the thermal CFT behavior. For example, $f$ parameterizes any quantity derivable from the free energy since $F = -\beta^{-1}\log\mathcal{Z}$, as well as the stress-tensor thermal one-point function since it too is proportional to a derivative of the partition function. 

    In order to check that our approximate partition function $\mathcal{Z}(\beta)$ can probe the high-temperature regime, we want to verify that $\mathcal{Z}(\beta)$ has behavior which approaches that of $\mathcal{Z}_{\text{high-}T}(\beta)$ in the $\beta\to 0$ limit. Practically speaking, this amounts to taking the high-temperature limit and extracting a value for $f$
    \begin{align}
        \lim_{\beta\to 0}\frac{\beta^2}{4\pi}\log\mathcal{Z}(\beta) &= f.
    \end{align}
    If our approximate partition function has converged in the high-temperature regime, then we expect that the value we extract for $f$ will be in agreement with previously reported values for the 3d Ising CFT. Literally taking this limit with our approximate CFT partition function with a finite number of operators will always yield zero. Since we are only approximately computing the partition function up to a cutoff $\Delta_\text{max}$ set by our knowledge of the CFT spectrum, the $\beta^2$ dependence will always eventually take over and force the limit to vanish.

    In order to actually extract a value for $f$ we will extrapolate from the region where we expect our partition function with a finite number of operators to be accurate, following a similar approach as \cite{Iliesiu:2018fao}. We first define a new function
    \begin{align}
        g_{\Delta_\text{max}}(\beta) &\equiv \frac{\beta^2}{4\pi}\log \mathcal{Z}_{\Delta_\text{max}}(\beta),
    \end{align}
    where $\mathcal{Z}_{\Delta_\text{max}}(\beta)$ is the sphere partition function up to the cutoff $\Delta_\text{max}$. Then, in order to approximate $f$, we will extrapolate $g_{\Delta_\text{max}}(\beta)$ out to $\beta = 0$ using data from a region where we are confident that $g_{\Delta_\text{max}}(\beta)$ has sufficiently converged to the full answer $g_\infty(\beta)$.

\begin{figure}[h!]
        \centering
        \includegraphics[width=0.9\linewidth]{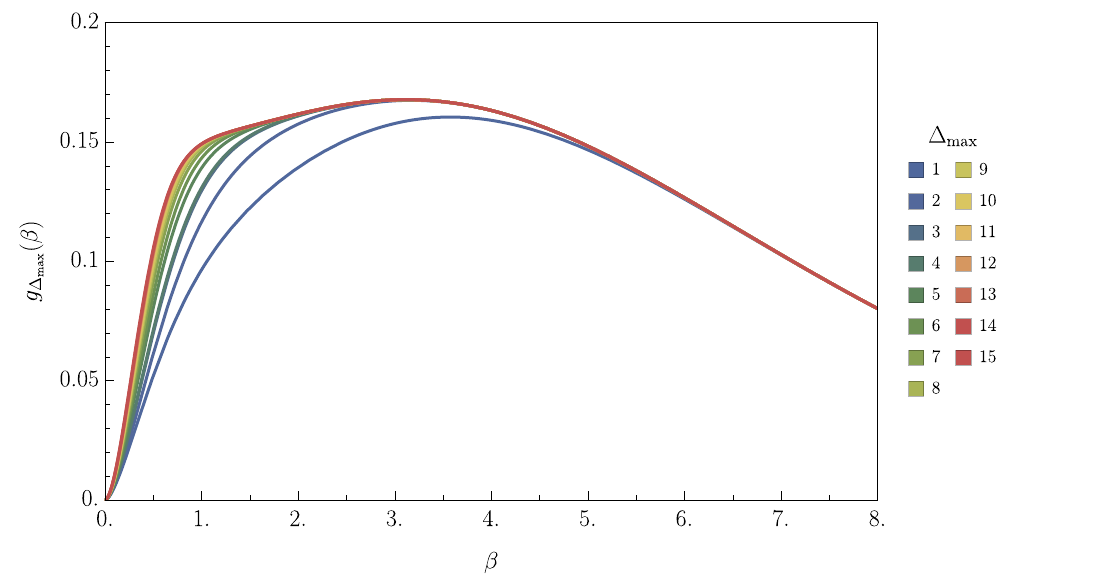}
        \caption{The family of curves $g_{\Delta_\text{max}}(\beta)$ plotted with an increasing number of primary operator contributions. $\Delta_\text{max}$ labels the highest dimension primary operator being included in each curve. The curves are color coded such that going from blue to red implies increasing $\Delta_\text{max}$. Implicitly all curves include the identity operator trivially.}
        \label{fig:gDeltaMaxByOp}
    \end{figure}

    The convergence of the $g_{\Delta_\text{max}}(\beta)$ to $g_\infty(\beta)$ is important for this whole procedure to be well-defined. In order to verify that the $g_{\Delta_\text{max}}(\beta)$ are indeed converging to some $g_\infty(\beta)$, we plot this family of curves for $\Delta_\text{max} = 1, 2, \ldots, 15$ in Fig. \ref{fig:gDeltaMaxByOp}. We observe that as $\Delta_\text{max}$ increases, the value of $\beta$ at which neighboring curves disagree is pushed to smaller $\beta$, in accordance with our expectation that $g_{\Delta_\text{max}}(\beta)$ more faithfully approximates $g_\infty(\beta)$ for increasing $\Delta_\text{max}$.

    We must now identify the smallest value of $\beta$ at which we can trust our best curve, i.e. the one with all operators included. We determine this smallest value of $\beta$ by comparing $g_{\Delta_\text{max}}(\beta)$ with different $\Delta_\text{max}$ and seeing where they differ. Comparison by eye shows us that the smallest value of $\beta$ we should trust for our extrapolation is
    \begin{align}
        \beta^* &\approx 1.4.
    \end{align}
    Finally, the form of our extrapolation is based on the thermal effective action, which tells us that as $\beta\to 0$ the function $g_\infty(\beta)$ behaves as
    \begin{align}
        g_\infty(\beta) &= f + c_2\beta^2 + c_4\beta^4 + c_6\beta^6 + c_8\beta^8 + \mathcal{O}(\beta^{10}),
    \end{align}
    and we have already identified the leading term with the free energy.

\begin{figure}[h!]
        \centering
        \includegraphics[width=0.9\linewidth]{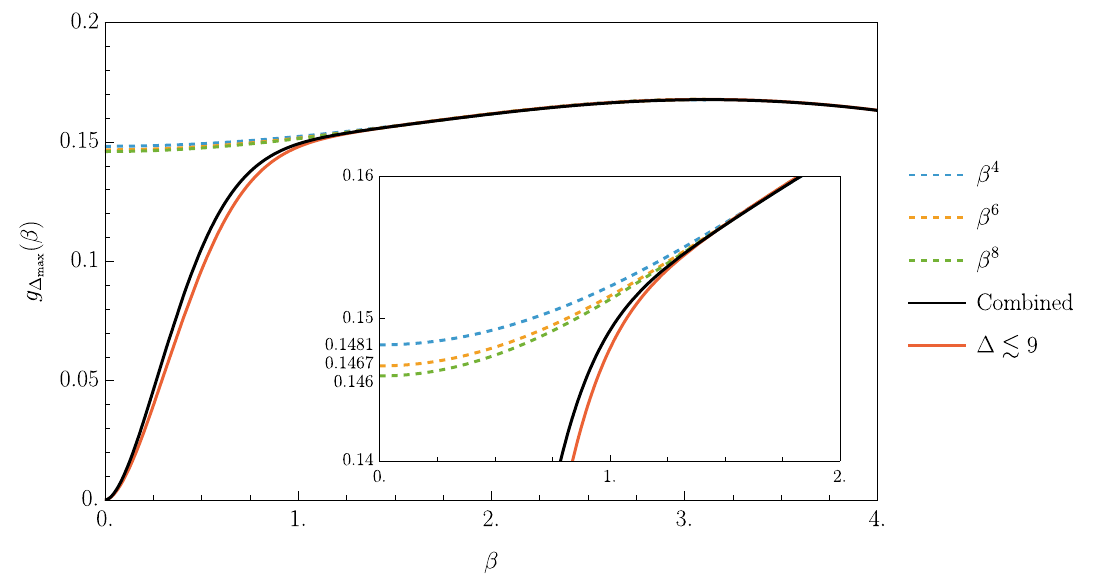}
        \caption{The curve $g_{\Delta_\text{max}}(\beta)$ with three different extrapolations out to $\beta = 0$. The black curve uses the combined $\Delta \lesssim 9$ and low spin $\Delta < 15$ (``combined'') CFT data, while the red curve uses only $\Delta \lesssim 9$. The dashed colored curves are for the extrapolations. The three extrapolations are even power series fits where the labels $\beta^4$, $\beta^6$, and $\beta^8$ indicate the highest power included in the fit. The inset blows up the region $0\leq \beta\leq 2$ within the window $[0.14,0.16]$ where the three extrapolations land.}
        \label{fig:gDeltaMaxExtrapolation}
    \end{figure}

    For our extrapolation we fit three different even power series of increasing maximum order ($\beta^4$, $\beta^6$, and $\beta^8$) over the window $\beta\in[1.4,3.11]$. The window's lower bound is set by $\beta^*$, while the upper bound is chosen to be the local maximum of $g_{\Delta_\text{max}}(\beta)$. We fix this upper bound since we expect that beyond this point the thermal physics is dominated by the lowest dimension primary operator $\sigma$. Our three extrapolations are plotted in Fig. \ref{fig:gDeltaMaxExtrapolation}. Additionally, we plot a $g_{\Delta_\text{max}}(\beta)$ restricted to the complete spectrum below $\Delta \lesssim 9$ to see if there is a noticeable advantage from the larger combined spectrum.

    We observe that the three different extrapolations all yield values for $f$ which agree down to two digits of precision. There is some sensitivity to the range of $\beta$ we use to do the fit. By looking at many of these different possible choices, we estimate that the first two digits are stable with the third digit spreading across a slightly wider range. We account for these equivalent choices by reporting that our method predicts the following value for $f$
    \begin{align}
        f &= 0.146(2),
    \end{align}
    where the error indicated (in the third digit) is intended to reflect the spread of predictions for $f$ given reasonable choices for the fitting window. Further, we observe that our larger spectrum does not noticeably affect the results. Both curves agree down to $\beta^*$ and do not differ until smaller $\beta$ where neither curve is valid.

    We compare this value of $f$ to other values in the literature in Table \ref{tab:fValsTable}. We find that our measured value for $f$ is consistent with the different Bootstrap studies and mildly inconsistent with what was found by Monte Carlo. 

     \begin{table}[h!]
        \centering
        \begin{tabular}{|c|C{2.1cm}|C{2.1cm}|C{2.1cm}|C{2.1cm}|}
            \hline
              & This work & MC \cite{PhysRevE.53.4414,PhysRevE.56.1642,PhysRevE.79.041142} & BS 2018 \cite{Iliesiu:2018zlz} & BS 2024 \cite{Barrat:2025wbi} \\
             \hline 
             $f$ & $0.146(2)$ & $0.153(1)$ & $0.1433(33)$ & $0.1432(51)$ \\
             \hline
        \end{tabular}
        \caption{Different measurements for $f$, the scheme-independent part of the free energy density, in the 3d Ising CFT. The values we quote come from previous Monte Carlo (MC) and Bootstrap (BS) studies.}
        \label{tab:fValsTable}
    \end{table}

    \subsection{The Thermal Expectation Value of \texorpdfstring{$\epsilon$}{epsilon} and \texorpdfstring{$b_\epsilon$}{beps}}

    Our  next goal is to compute the thermal one-point function of $\epsilon$ on the sphere, $\langle\epsilon\rangle_{S^1_\beta\times S^2}$. Recall that the thermal one-point function of a scalar primary operator on the sphere can be computed as a sum over OPE coefficients with Boltzmann weights
    \begin{align}
        \langle \phi \rangle_{S^1_\beta\times S^2} &= \frac{1}{\mathcal{Z}(\beta)}\sum_{\Delta} \lambda_{\phi\mathcal{OO}}e^{-\beta\Delta},\label{eq:1ptFuncGeneralSum}
    \end{align}
    where $\mathcal{Z}(\beta)$ is the partition function on the sphere as defined in (\ref{eq:PartitionFuncChiDefn}) and $\Delta$ goes over all states in the CFT, both primaries and descendants. As with the partition function example, we will be working with a limited pool of data to compute the one-point function. Thus, in order to get as much mileage out of the data as possible, we will reorganize the sum over states to run over primary operators only 
    \begin{align}
        \langle\phi\rangle_{S^1_\beta\times S^2} &= \frac{1}{\mathcal{Z}(\beta)}\sum_{\Delta\text{ primary}}\sum_{a=0}^{J}\lambda^{a}_{\phi\mathcal{OO}}g^{\Delta_\phi,a}_{\Delta,J}(\beta),\label{eq:1ptFuncBlockSum}
    \end{align}
    where $g^{\Delta_\phi,a}_{\Delta,J}(\beta)$ is a thermal conformal block for a scalar $\phi$ with external states $\mathcal{O}$ of dimension $\Delta$ and spin $J$, $\lambda^{a}_{\phi\mathcal{OO}}$ is an OPE coefficient, and $a$ labels three-point function tensor structures. Finally, from Section \ref{sec:FiniteTCFTs} we know that the scalar thermal one-point function on the sphere has the specific closed-form 
    \begin{align}
        \langle \phi \rangle_{S^1_\beta\times S^2} &= \frac{h(\beta)}{\beta^{\Delta_\phi}},\label{eq:Sph1ptFuncExact}
    \end{align}
    with $h(\beta)$ an unknown function that recovers the flat-space limit at $\beta=0$:
    \begin{align}
        \langle \phi \rangle_{S^1_\beta\times S^2} \stackrel{\beta\sim 0}{\sim}
        \frac{b_\phi}{\beta^{\Delta_\phi}},\label{eq:Flat1ptFuncExact}
    \end{align}
    where $b_\phi$ is the thermal vev for this operator.

    Before we can construct the thermal one-point function of $\epsilon$ we must deal with the fact that unlike the characters, the thermal blocks $g^{\Delta_\phi,a}_{\Delta,J}(\beta)$ do not generically have known closed-form solutions.\footnote{It was originally found in \cite{Gobeil:2018fzy} that there is a closed-form solution when the external state $\mathcal{O}$ is also a scalar. More recently in \cite{Buric:2026pes}, more results for certain closed-forms of thermal blocks with spinning external states have also been derived. The authors of \cite{Buric:2026pes} further claim that it is possible to derive closed-forms for any of the scalar thermal blocks at vanishing chemical potential, but so far on a case-by-case basis alone.} Thus, we must make sure to approximate the thermal blocks in such a way that we saturate all descendant contributions in the regime of $\beta$ that we can probe. Since the thermal blocks for scalar external states have the following known closed-form solution \cite{Gobeil:2018fzy}
    \begin{align}
        \frac{q^{\Delta}}{(1-q)^{2\Delta}}{}_3F_2\left(\Delta - 1,\Delta - \frac{\Delta_\epsilon}{2}, -\frac{3}{2} + \frac{\Delta_\epsilon}{2} + \Delta; \Delta, 2\Delta - 2; -\frac{4q}{(q-1)^2} \right),
    \end{align}
    we use it in our calculation, where $q=e^{-\beta}$. For general spinning thermal blocks for arbitrary tensor structures, we compute them as a low-temperature expansion in $q$ using the code from \cite{Buric:2024kxo}. We generate all spinning thermal blocks as expansions up through $\mathcal{O}(q^{14})$. This expansion captures all descendants through level 14, which should be enough to properly populate the CFT Hilbert space down to any value of $\beta$ we could hope to probe given our cutoff in $\Delta$.

\begin{figure}[h!]
        \centering
        \includegraphics[width=0.9\linewidth]{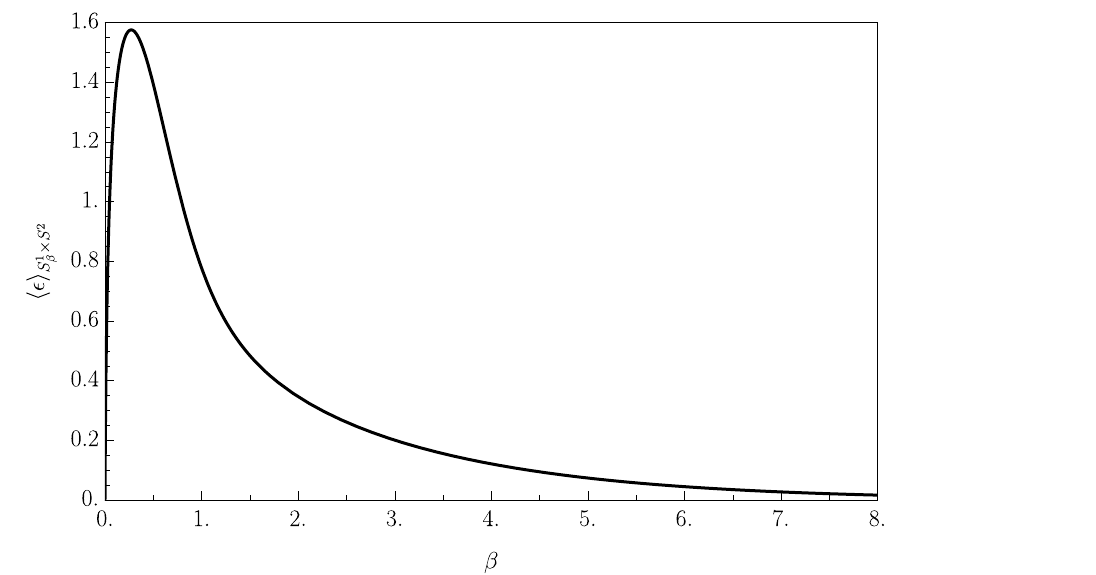}
        \caption{The approximate thermal one-point function of $\epsilon$ on the sphere, $\langle\epsilon\rangle_{S^1_\beta\times S^2}$. The one-point function is constructed using all operators below $\Delta \lesssim 9$ and all operators below $\Delta < 15$ with $J\leq 4$ in the 3d Ising CFT. The curve incorrectly vanishes as $\beta\to 0$ rather than diverging as $\beta^{-\Delta_\epsilon}$.}
        \label{fig:EpsilonThermal1ptFunc}
    \end{figure}

    To construct $\langle\epsilon\rangle_{S^1_\beta\times S^2}$, we use all primary operators below $\Delta \lesssim 9$ and all primary operators below $\Delta < 15$ with $J \leq 4$. The result is plotted in Fig. \ref{fig:EpsilonThermal1ptFunc}. The curve begins to rise sharply at small $\beta$, as expected from the flat-space behavior $\sim \beta^{-\Delta_\epsilon}$, before turning around and dropping to zero, as expected from our finite set of OPE data. 

 To extract the small $\beta$ behavior, we again will perform a fit over the range we trust our calculation of $\langle\epsilon\rangle_{S^1_\beta\times S^2}$, and extrapolate to $\beta \sim 0$, as we did for the partition function in Section \ref{sec:PartFunc}. Define
    \begin{align}
        h_{\Delta_\text{max}}(\beta) &\equiv \beta^{\Delta_\epsilon}\langle \epsilon \rangle_{S^1_\beta\times S^2}^{\Delta_\text{max}}, 
    \end{align}
    where $\langle\epsilon\rangle_{S^1_\beta\times S^2}^{\Delta_\text{max}}$ is the thermal one-point function of $\epsilon$ on the sphere up to the cutoff $\Delta_\text{max}$. For each value of $\Delta_{\rm max}$, we impose the same cutoff on the partition function (which enters when we divide by it to normalize the one-point function) as we do on the sum over OPE coefficients, meaning that $\langle\epsilon\rangle_{S^1_\beta\times S^2}^{\Delta_\text{max}}$ is normalized by $\mathcal{Z}_{\Delta_\text{max}}(\beta)$. In order to approximate $b_\epsilon$, we will again extrapolate $h_{\Delta_\text{max}}(\beta)$ out to $\beta = 0$ using data from a region where we are confident that our approximate function has converged to the exact result, $h_\infty(\beta)$.

    \begin{figure}[h!]
        \centering
        \includegraphics[width=0.9\linewidth]{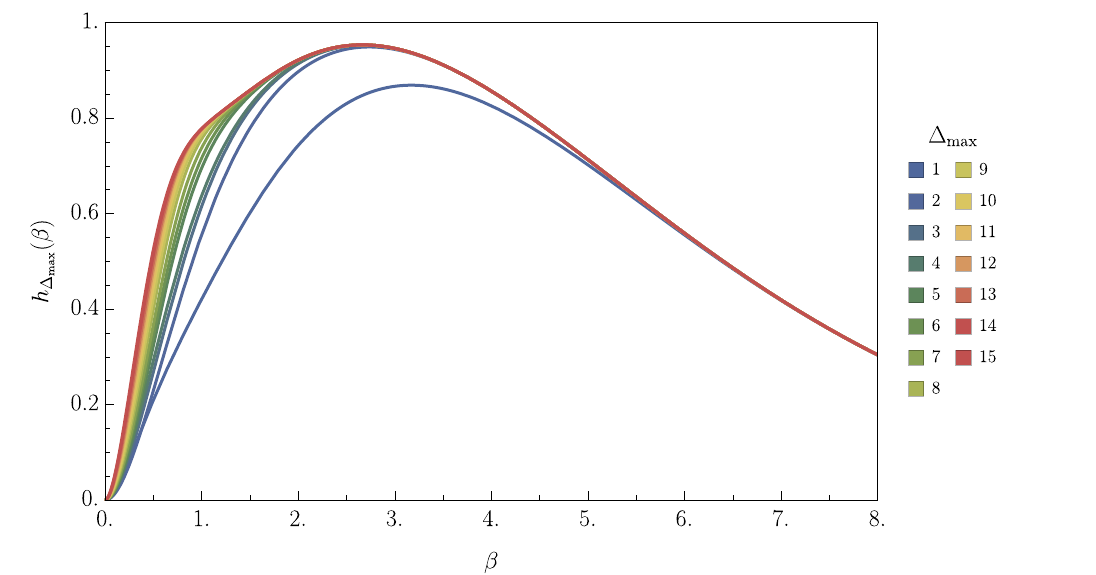}
        \caption{The family of curves $h_{\Delta_\text{max}}(\beta)$ plotted with an increasing number of primary operator contributions. $\Delta_\text{max}$ labels the highest dimension primary operator being included in each curve. The curves are color coded such that going from blue to red implies increasing $\Delta_\text{max}$.}
        \label{fig:hDeltaMaxByOp}
    \end{figure}
    
    To verify that the $h_{\Delta_\text{max}}(\beta)$ are converging to some $h(\beta)$, we plot the family of curves $h_{\Delta_\text{max}}(\beta)$ for increasing $\Delta_\text{max}$ in Fig. \ref{fig:hDeltaMaxByOp}. We can clearly see that as $\Delta_\text{max}$ increases, the curves agree to smaller and smaller values of $\beta$, as expected. We can now identify the smallest value of $\beta$ that we can trust our best $h_{\Delta_\text{max}}(\beta)$. By again comparing when curves of different $\Delta_\text{max}$ differ, we conservatively set
    \begin{align}
        \beta^* &\approx 1.2
    \end{align}
    as the smallest value of $\beta$ that we will trust for our extrapolation. Finally, our extrapolation of $h_{\Delta_\text{max}}(\beta)$ relies on $h(\beta)$ being well-described by a thermal effective action as $\beta\to 0$:
    \begin{align}
        h(\beta) &= b_\epsilon + b_2 \beta^2 + b_4 \beta^4 + b_6\beta^6 + b_8\beta^8 + \mathcal{O}(\beta^{10}),
    \end{align}
    where the leading term is fixed by the boundary condition $h(0) = b_\epsilon$.

\begin{figure}[h!]
        \centering
        \includegraphics[width=0.9\linewidth]{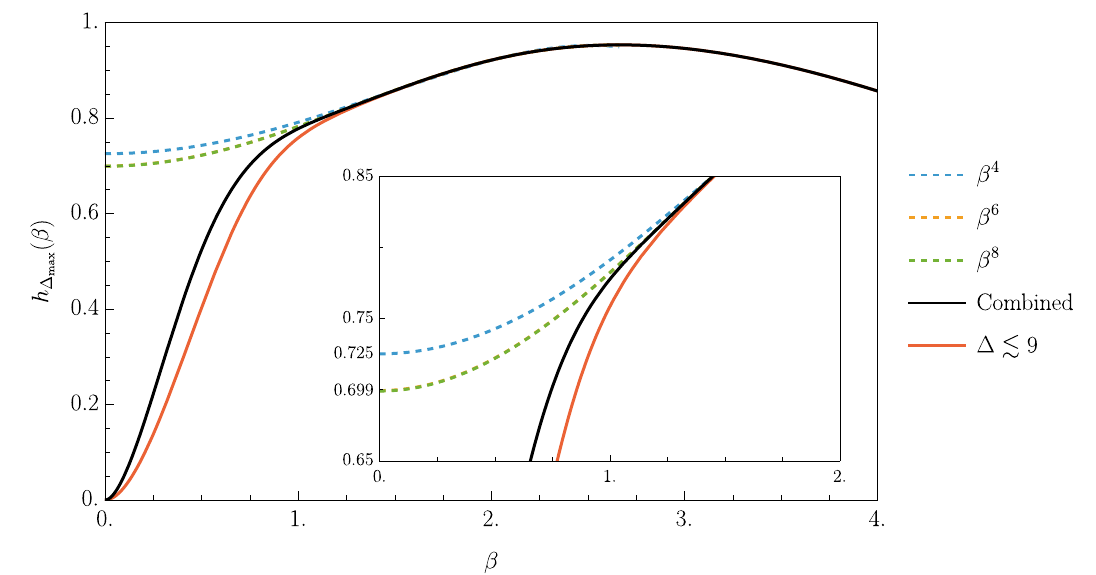}
        \caption{The curve $h_{\Delta_\text{max}}(\beta)$ with three different extrapolations out to $\beta = 0$. The black curve uses the combined $\Delta \lesssim 9$ and low spin $\Delta < 15$ (``combined'') CFT data, while the red curve uses only $\Delta \lesssim 9$. The dashed colored curves are for the extrapolations. The three extrapolations are even power series fits where the labels $\beta^4$, $\beta^6$, and $\beta^8$ indicate the highest power included in the fit. The inset blows up the region $0\leq \beta\leq 2$ within the window $[0.65,0.85]$ where the three extrapolations land. Note that the $\beta^6$ and $\beta^8$ curves are overlapping.}
        \label{fig:hDeltaMaxExtrapolation}
    \end{figure}

    We now perform our extrapolation where we fit three different even power series of increasing maximum order ($\beta^4$, $\beta^6$, and $\beta^8$) over the window $\beta\in[1.2,2.663]$. The window's lower bound is set by $\beta^*$, while the upper bound is chosen to be the local maximum of $h_{\Delta_\text{max}}(\beta)$. Just as in the partition function case, we expect that beyond this upper bound the thermal physics is dominated by just the lowest dimension primary operator $\sigma$. We plot our three extrapolations with $h_{\Delta_\text{max}}(\beta)$ in Fig. \ref{fig:hDeltaMaxExtrapolation}. Additionally, we plot an $h_{\Delta_\text{max}}(\beta)$ restricted to the complete spectrum below $\Delta \lesssim 9$ to see if there is a noticeable advantage from our larger spectrum.

    We observe that the three extrapolations yield values of $b_\epsilon$ which agree down to one digit of precision. Since our fitting window is not the only valid choice, we again scan over fits with adjusted fitting windows and report that
    our method predicts the following value for $b_\epsilon$
    \begin{align}
        b_\epsilon = 0.70(2),
    \end{align}
    where we emphasize that the reported uncertainty is only accounting for the spread of predictions of $b_\epsilon$ given reasonable choices for the fitting window. We note that our larger combined spectrum has a much more noticeable effect by yielding a smaller $\beta^*$ than the pure $\Delta \lesssim 9$ curve. However, fitting in the trusted region of the $\Delta \lesssim 9$ curve does not drastically alter the extrapolated value for $b_\epsilon$.

    Now that we have a prediction for $b_\epsilon$, we can compare with the other values in the literature. The full set of literature values for $b_\epsilon$ have been collected in Table \ref{tab:bEpsilonValsTable}. We find that our measurement is consistent with both Bootstrap values. Given these results, we can conclude that our direct construction of the thermal one-point function of $\epsilon$ on the sphere has enabled us to probe the high-temperature regime.
    
    Perhaps surprisingly, given the large deviation we found from the leading order ETH prediction with the OPE coefficients at $\Delta \lesssim 9$, it turns out that including only $\Delta \lesssim 9$ operators in the Boltzmann sum produces essentially the same value of $b_\epsilon$. In other words, even though their OPE coefficients are significantly higher in this range than ETH predicts, those same OPE coefficients in the macrocanonical ensemble match the same quantity $b_\epsilon$ much more accurately. We are therefore motivated to see if we can do better in the opposite direction, from thermal expectation values to OPE coefficients, by including corrections to the ETH prediction. In the next section we will return to ETH and compute many subleading terms to see if we can make contact with the low-dimension physics from the high-temperature side.

    \begin{table}[h!]
        \centering
        \begin{tabular}{|c|C{2.1cm}|C{2.1cm}|C{2.1cm}|}
            \hline
              & This work & BS 2018 \cite{Iliesiu:2018zlz} & BS 2024 \cite{Barrat:2025wbi} \\
             \hline 
             $b_\epsilon$ & $0.70(2)$ & $0.63(7)$ & $0.71(14)$ \\
             \hline
        \end{tabular}
        \caption{Different measurements for $b_\epsilon$, the thermal vev of $\epsilon$, in the 3d Ising CFT. The values we quote come from previous Bootstrap (BS) studies.}
        \label{tab:bEpsilonValsTable}
    \end{table}

\section{Corrections to ETH}\label{sec:ETHCorrections}

    We now explore whether ETH can make contact with low-dimension CFT data by computing many subleading corrections. Although many types of corrections may be organized by and incorporated through the thermal effective action, we are specifically interested in those corrections which are universal (i.e. they depend on $f$ and $b_\epsilon$ alone), since such corrections are calculable.  If we are fortunate,  they will also be large relative to theory-dependent corrections. We saw an example of this already in Section \ref{sec:FSandETH} with the first universal subleading correction to ETH (\ref{eq:ETHwithSubleading}), where its coefficient is $\mathcal{O}(1)$ and the dominant term until $\Delta\approx 5$.
    
    Thus, our goal is to compute a tower of these subleading universal corrections to ETH and compare with the low-dimension OPE data. With the tower of computed corrections in hand, we will attempt to resum the series of universal terms to see if this provides further improvements which allows the high-temperature expansion to make contact with the low-dimension data.

    In order to compute corrections to ETH, we will make use of the inversion formula machinery developed in \cite{Buric:2025uqt}. As a warm-up, we will look at the density of primary states $\rho(\Delta,J)$ and the calculation of its infinite tower of universal corrections. We will then resum a finite number of these corrections using Borel-Pad\'e resummation. After, we will apply the same ideas to the calculation of universal subleading corrections to ETH and compare directly with the low-dimension OPE data. 

    In what follows we must turn on a chemical potential $\mu$ for angular momentum. Thus, objects like the thermal conformal blocks can in general depend on $\beta$, $\mu$, and $\theta$ the polar coordinate on $S^2$. At points it may be more convenient to write objects in terms of the variables $(q,y,s)$
    \begin{align}
        q &= e^{-\beta}, & y &= e^{i\mu}, & s &= \sin^2\theta.
    \end{align}
    In this section we also denote the thermal one-point function on the sphere at non-zero chemical potential by $\langle\phi\rangle_{\beta,\mu}$ in order to simplify notation. Finally, the upcoming calculations and figures depend on the parameters $f$ and $b_\epsilon$. For this section, we choose to use the values we extracted in Section \ref{sec:FiniteTObservablesOnSphere}
    \begin{align}
        b_\epsilon &\approx 0.70, & f &\approx 0.146.
    \end{align}
    
    \subsection{Warm-Up: The Density of Primary States}\label{sec:DensityOfPrimaryStates}

    We want to compute the high energy density of states for a general 3d CFT. By first approximating the dimensions of primary operators as continuous, we can decompose the partition function on the sphere as 
    \begin{align}
        \mathcal{Z}(\beta,\mu) &= \int_0^\infty d\Delta \sum_{J=0}^\infty \rho(\Delta,J)\chi_{\Delta,J}(\beta,\mu),
    \end{align}
    where $\chi_{\Delta,J}(\beta,\mu)$ is the chemical potential dependent character
    \begin{align}
        \chi_{\Delta,J}(q,y) &= \frac{q^{\Delta}y^{-J}(1-y^{(2J+1)})}{(1-q)(1-y)(1-qy)(1-qy^{-1})}.
    \end{align}
    Using the character inversion formula from \cite{Buric:2025uqt} allows us to express the density of primary states as 
    \begin{align}
        \rho(\Delta,J) &= \frac{1}{2\pi i}\int_{\gamma-i\infty}^{\gamma+i\infty}d\beta \int_{-\pi}^{\pi}d\mu\,\omega(\beta,\mu)\chi_{3-\Delta,J}(\beta,\mu)\mathcal{Z}(\beta,\mu),\label{eq:DensityOfStatesDef}
    \end{align}
    where the measure is orthogonal on the space of characters and takes the form
    \begin{align}
        \omega(\beta,\mu) &= \left(\frac{8}{\sqrt{\pi}}\sinh\left(\frac{\beta}{2}\right)\sin\left(\frac{\mu}{2}\right)\sinh\left(\frac{\beta+i\mu}{2}\right)\sinh\left(\frac{\beta-i\mu}{2}\right)\right)^2.\label{eq:MeasureForChi}
    \end{align}
    Since we are interested in the universal behavior of the high energy density of states, we take the partition function to be the leading term from the thermal effective action 
    \begin{align}
        \mathcal{Z}(\beta,\mu) &\approx e^{\frac{4\pi f}{\beta^2+\mu^2}}.
    \end{align}
    We will proceed by evaluating (\ref{eq:DensityOfStatesDef}) via saddle point approximation. Computing the integrals this way generates $\rho(\Delta,J)$ as a series expansion in $\Delta^{-1/3}$, which has been done in both \cite{Benjamin:2023qsc} and \cite{Buric:2025uqt}. We will review the procedure and extend the calculation to higher orders, eventually attempting to resum the series using Borel-Pad\'e resummation in order to extend the validity of the series to smaller $\Delta$. 
    
    Starting with the saddle point approximation, it is useful to first change coordinates
    \begin{align}
        \beta &= \frac{\beta_L + \beta_R}{2}, & \mu &= \frac{\beta_L - \beta_R}{2i},\label{eq:betaLbetaR}
    \end{align}
    which allows us to rewrite (\ref{eq:DensityOfStatesDef}) as 
    \begin{align}
        \rho(\Delta,J) &= -\frac{1}{4\pi}\iint d\beta_L d\beta_R\,G(\beta_L,\beta_R)e^{\frac{4\pi f}{\beta_L\beta_R}+\frac{\beta_L+\beta_R}{2}(\Delta-3)},
    \end{align}
    where $\beta_L,\beta_R$ run over $(\gamma-i\infty,\gamma+i\infty)$ and we have defined a $\Delta$-independent piece
    \begin{align}
        G(\beta_L,\beta_R) &\equiv \omega(\beta_L,\beta_R)\hat{\chi}_{3-\Delta,J}(\beta_L,\beta_R),
    \end{align}
    by extracting the $\Delta$ dependence from the shadow character by defining 
    \begin{align}
        \hat{\chi}_{3-\Delta,J}(\beta_L,\beta_R) &\equiv \chi_{3-\Delta,J}(\beta_L,\beta_R)e^{-\frac{\beta_L+\beta_R}{2}(\Delta-3)}.
    \end{align}
    Note that when performing this change of coordinates we have also extended the domain of integration, generating a non-perturbative piece. In the $\Delta\to\infty$ limit where we will perform the saddle point analysis, this piece is suppressed.

    We can now carry out the saddle point procedure. By defining 
    \begin{align}
        \mathcal{F}(\beta_L,\beta_R) &\equiv -\left(\frac{4\pi f}{\beta_L\beta_R}+\frac{\beta_L+\beta_R}{2}\Delta\right),
    \end{align}
    it is easy to extract that the saddle occurs at
    \begin{align}
        \beta_L^* &= \beta_R^* = \left(\frac{8\pi f}{\Delta}\right)^{1/3},
    \end{align}
    noting that at this point we are taking $\Delta\to \infty$ and $J = \text{fixed}$. However, we will follow along with \cite{Buric:2025uqt} for consistency and first perform one more change of variable
    \begin{align}
        \beta_L &= \left(\frac{8\pi f}{\Delta}\right)^{1/3}\left(1+ix_L\Delta^{-1/3}\right), & \beta_R &= \left(\frac{8\pi f}{\Delta}\right)^{1/3}\left(1+ix_R\Delta^{-1/3}\right).\label{eq:xLxR}
    \end{align}
    Thus, the integral we are now left with is 
    \begin{align}
        \frac{(8\pi f)^{2/3}}{4\pi\Delta^{4/3}}e^{3\pi^{1/3}f^{1/3}\Delta^{2/3}}\iint dx_L dx_R\,G(x_L,x_R)e^{-\pi^{1/3}f^{1/3}(x_L^2+x_L x_R + x_R^2)-\sum_{k>2}\widetilde{\mathcal{F}}^{(k)}(x_L,x_R)},
    \end{align}
    where $x_L$, $x_R$ run over $(-\infty,\infty)$ and the $\widetilde{\mathcal{F}}^{(k)}$ come from expanding 
    \begin{align}
        \widetilde{\mathcal{F}} &= \mathcal{F}(x_L,x_R) + \frac{3}{2}\left[ \left(\frac{8\pi f}{\Delta}\right)^{1/3}\left(1+ix_L\Delta^{-1/3}\right) + \left(\frac{8\pi f}{\Delta}\right)^{1/3}\left(1+ix_R\Delta^{-1/3}\right)\right]
    \end{align}
    around the saddle point $x_L^* = x_R^* = 0$ beyond quadratic order. For example, the cubic and quartic terms are 
    \begin{align}
        \widetilde{\mathcal{F}}^{(3)}(x_L,x_R) &= -\frac{if^{1/3}\pi^{1/3}(6i + x_L^3 + x_L^2 x_R + x_L x_R^2 + x_R^3)}{\Delta^{1/3}}, \\
        \widetilde{\mathcal{F}}^{(4)}(x_L,x_R) &= -\frac{f^{1/3}\pi^{1/3}(x_L^4 + x_L^3 x_R + x_L^2 x_R^2 + x_L(x_R^3 - 3i) + x_R(x_R^3 - 3i))}{\Delta^{2/3}}.
    \end{align}
    This setup allows us to carry out the saddle point analysis to very high order. For general $J$ we generated the series up to $\mathcal{O}(\Delta^{-7})$ relative to the leading order term, which is equivalent to saying that we generated 22 terms of the series. Below are the first few terms including the first $J$-dependent term
    \begin{align}
        \rho(\Delta,J) &= \frac{8\pi^{2/3}f^{5/3}(2J+1)}{\sqrt{3}\Delta^{11/3}}e^{3\pi^{1/3}f^{1/3}\Delta^{2/3}}\Bigg[1 - \frac{3\pi^{1/3}f^{1/3}}{\Delta^{1/3}} + \frac{5(9\pi f - 7)}{9\pi^{1/3}f^{1/3}\Delta^{2/3}} \label{eq:DenOfStatesSeries}\\
         &+ \frac{103-36\pi f}{6\Delta} + \frac{5600 + 4644\pi^2f^2 - 45\pi f(18J^2 + 18J + 713)}{810 f^{2/3}\pi^{2/3}\Delta^{4/3}} + \mathcal{O}(\Delta^{-5/3})\Bigg].\nonumber
    \end{align}
    Note that higher order terms have been omitted for space.

\begin{figure}[h!]
        \centering
        \includegraphics[width=0.9\linewidth]{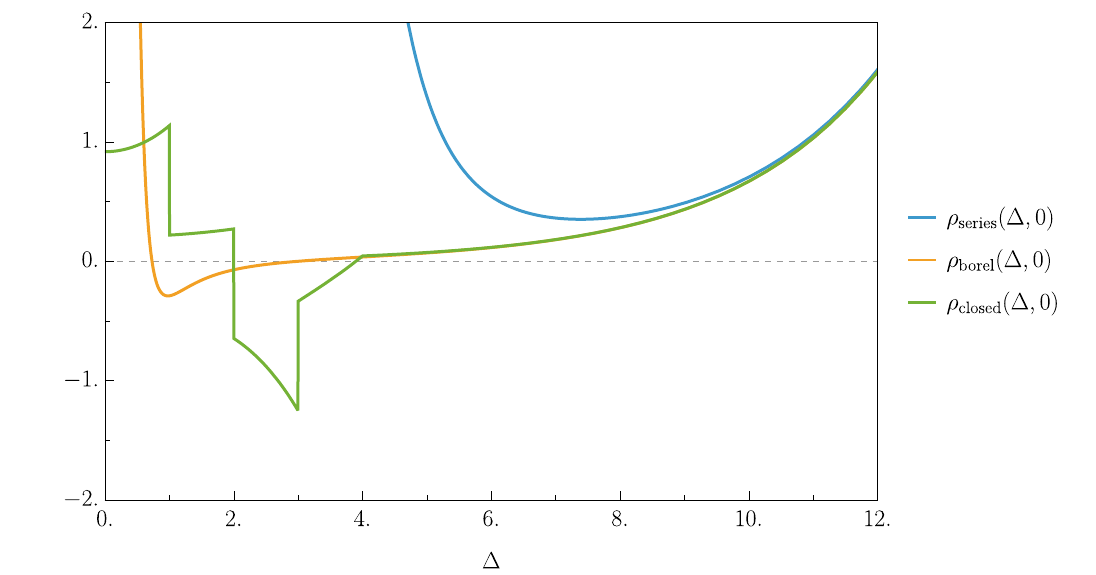}
        \caption{Comparison of the spin-0 density of primary states as: a 22 term asymptotic series, a Borel-Pad\'e resummation, and a closed-form expression.}
        \label{fig:rhoClosedForm}
    \end{figure}

     In order to extend the range of this formula, at least approximately, we also attempt to resum the perturbative expansion (\ref{eq:DenOfStatesSeries}) using Borel-Pad\'e resummation (see Appendix \ref{app:BorelPade} for details). Taking $J = 0$ as a simple example, we compare the raw 22 term series with the Borel-Pad\'e resummation in Fig. \ref{fig:rhoClosedForm}. We observe that this Borel-Pad\'e resummation is finite and well-behaved down to $\Delta \approx 4$, with unphysical behavior taking over at $\Delta < 4$. For sufficiently large $\Delta$, the raw series and resummation agree.

    It is not obvious how seriously we should take the Borel-Pad\'e resummation. To alleviate this concern, we notice that the density of primary states is special and the integral (\ref{eq:DensityOfStatesDef}) is amenable to slightly more analytic work that allows us to compute a closed-form expression (\ref{eq:DenOfStatesClosedForm}). In particular, this closed-form expression resums the entire perturbative series. By comparing this exact resummation with Borel-Pad\'e in Fig. \ref{fig:rhoClosedForm}, we observe that both exactly agree down to $\Delta = 4$ for $J = 0$. This behavior is mirrored for $J>0$ with the threshold $\Delta$ increasing with $J$. This provides us with confidence that Borel-Pad\'e resumming these kinds of expansions does allow us to extract as much mileage as is available.

    Before going on to ETH, in the next section we go through the derivation of the closed-form expression for the density of primary states (\ref{eq:DenOfStatesClosedForm}). Then, as an application, we integrate this closed-form expression to get a prediction for the count of CFT states, $N(\Delta,J)$, and compare it with data obtained from the Fuzzy Sphere. The details in the next subsection have no bearing on the ETH results to come, and the ETH motivated reader can skip directly to Section \ref{sec:UniversalSubleadingETH}.

    \subsection{Closed-Form Density of Primary States and Borel Resummation}\label{sec:RhoExactBonus}

    As previously mentioned, it turns out that the density of primary states is special and more can be done beyond the perturbative expansion and subsequent Borel-Pad\'e resummation. With a simple approximation, we can evaluate the integral in closed-form, which in turn resums the entire perturbative expansion.

    Returning to the original integral (\ref{eq:DensityOfStatesDef}), we first expand the partition function as an infinite series and then swap the order of summation and integration
   \begin{align}
        \rho(\Delta,J) &= \sum_{n=0}^{\infty}\frac{(4\pi f)^n}{n!}\frac{1}{2\pi i}\int_{\gamma-i\infty}^{\gamma+i\infty}d\beta \int_{-\pi}^{\pi}d\mu\,\omega(\beta,\mu)\chi_{3-\Delta,J}(\beta,\mu)\frac{1}{(\beta^2+\mu^2)^n}. \label{eq:rhoIntExpExpanded}
    \end{align}
    Then, we approximate the $\mu$-integral by extending the bounds from $[-\pi,\pi]\to(-\infty,\infty)$, allowing us to recast the $\mu$-integral as a contour integral that makes the evaluation trivial. Finally, we integrate term-by-term and resum the result to arrive at a closed-form expression. 
    
    An important caveat to address is the possibility of $\mathcal{O}(1)$ errors at small $\Delta$. This concern can be addressed directly by using the exact $[-\pi,\pi]$ bounds of the $\mu$-integral and carrying out a more arduous semi-analytic calculation, the details of which are in Appendix \ref{app:rhoExact}. By comparing the closed-form expression (\ref{eq:DenOfStatesClosedForm}) that we will derive to the more precise calculation in Appendix \ref{app:rhoExact}, one finds qualitatively similar results for $\Delta\lesssim 4$ and quantitatively similar results for $\Delta \gtrsim 4$. 

    Before we go on and evaluate the integral (\ref{eq:rhoIntExpExpanded}) for the two cases $n=0$ and $n>0$, it is first helpful to simplify the product of the measure and the shadow character to the following simple sums
    \begin{align}
        \frac{1}{4\pi}\sum_{s=\pm 1}\left[\sum_{\delta\in\{0,3\}}\sum_{j\in\{0,1\}} - \sum_{\delta\in\{1,2\}}\sum_{j\in\{-1,2\}}\right](-1)^{\delta+j}e^{\beta(\Delta-\delta)}e^{i\mu s(J+j)}.\label{eq:simplifiedRhoIntegrand}
    \end{align}
    This simplification makes it manifest that the integral's general structure is of the form
    \begin{align}
        \frac{1}{2\pi i}\int_{\gamma-i\infty}^{\gamma+i\infty}d\beta\int d\mu\,\frac{e^{\beta \overline{\Delta}}e^{i\mu\overline{J}}}{(\beta^2+\mu^2)^n},
    \end{align}
    where $\overline{\Delta} = \Delta - \delta$ and $\overline{J} = s(J + j)$.

    We start by tackling the trivial $n=0$ case. By using the original $\mu\in[-\pi,\pi]$ bounds, we can immediately show that 
    \begin{align}
        \frac{1}{2\pi i}\int_{\gamma-i\infty}^{\gamma+i\infty}d\beta\int_{-\pi}^{\pi} d\mu\,e^{\beta \overline{\Delta}}e^{i\mu\overline{J}} &= 2\pi\delta_{\overline{{J}},0}\delta(\overline{\Delta}).
    \end{align}
    Generalizing now to $n>0$, we can first show that the integral over $\mu$ yields 
    \begin{align}
        \int_{-\infty}^{\infty}d\mu\,\frac{e^{i\mu \overline{J}}}{(\beta^2+\mu^2)^n} &= \frac{\sqrt{\pi}}{2^{n-\frac{3}{2}}\Gamma(n)}\left(\frac{\lvert\overline{J}\rvert}{\beta}\right)^{n-\frac{1}{2}}K_{n-\frac{1}{2}}\left(\beta\lvert\overline{J}\rvert\right),
    \end{align}
    where $K_\alpha(x)$ is a modified Bessel function of the second kind. We can then move on to the final integral, which can again be carried out by contour integration 
    \begin{align}
        \frac{1}{2\pi i}\int_{\gamma-i\infty}^{\gamma+i\infty}d\beta\,\left(\frac{\lvert\overline{J}\rvert}{\beta}\right)^{n-\frac{1}{2}}K_{n-\frac{1}{2}}\left(\beta\lvert\overline{J}\rvert\right)e^{\beta\overline{\Delta}} &= \frac{\sqrt{\pi}}{2^{n-\frac{1}{2}}\Gamma(n)}\left(\overline{\Delta}^2-\lvert\overline{J}\rvert^2\right)^{n-1}\theta(\overline{\Delta}-\lvert\overline{J}\rvert).
    \end{align}
    Putting everything together, we have shown that 
    \begin{align}
        \frac{1}{2\pi i}\int_{\gamma-i\infty}^{\gamma+i\infty}d\beta\int_{-\infty}^{\infty} d\mu\,\frac{e^{\beta \overline{\Delta}}e^{i\mu\overline{J}}}{(\beta^2+\mu^2)^n} &= \frac{4\pi}{4^n\Gamma(n)^2}\left(\overline{\Delta}^2-\lvert\overline{J}\rvert^2\right)^{n-1}\theta\left(\overline{\Delta} - \lvert\overline{J}\rvert\right).
    \end{align}
    The final step is to take care of the sum over $n$. Fortuitously, this sum can be carried out exactly and written as a generalized Hypergeometric function 
    \begin{align}
        \sum_{n=1}^{\infty}\frac{(4\pi f)^n}{n!}\frac{\left(\overline{\Delta}^2-\lvert\overline{J}\rvert^2\right)^{n-1}}{4^n\Gamma(n)^2} &= \pi f{}_0F_2\left(;1,2;\pi f(\overline{\Delta}^2-\lvert\overline{J}\rvert^2)\right).
    \end{align}
    Therefore, the density of primary states can be written as 
    \begin{align}
        \rho_\text{closed}(\Delta,J) &= \hat{\rho}_0(\Delta,J) + \left[\sum_{\delta\in\{0,3\}}\sum_{j\in\{0,1\}} - \sum_{\delta\in\{1,2\}}\sum_{j\in\{-1,2\}}\right](-1)^{\delta+j}\hat{\rho}(\Delta-\delta,J+j),\label{eq:DenOfStatesClosedForm}
    \end{align}
    where we have defined
    \begin{align}
        \hat{\rho}_0(\Delta,J) &\equiv \delta_{J,0}[\delta(\Delta) - \delta(\Delta-3)] - \delta_{J,1}[\delta(\Delta-1) - \delta(\Delta-2)],\label{eq:rhoHatNaught} \\
        \hat{\rho}(\Delta,J) &\equiv 2\pi f{}_0F_2\left(;1,2;\pi f(\Delta^2-\lvert J\rvert^2)\right)\theta(\Delta - \lvert J \rvert),\label{eq:RhoClosedFormSeed}
    \end{align}
    and it should be understood that $\hat{\rho}_0(\Delta,J)$ comes from the $n=0$ pieces.

    It is interesting to briefly interpret $\hat{\rho}_0(\Delta,J)$. Physically, we can understand $\hat{\rho}_0(\Delta,J)$ as capturing the vacuum contribution to the partition function which is trivially always present since $\mathcal{Z}\supset 1$. Given that we can decompose this contribution into characters 
    \begin{align}
        1 &= \chi_{0,0}(\beta,\mu) - \chi_{3,0}(\beta,\mu) - \chi_{1,1}(\beta,\mu) + \chi_{2,1}(\beta,\mu),
    \end{align}
    and (\ref{eq:DensityOfStatesDef}) is orthogonal on the space of characters, it is no surprise that $\hat{\rho}_0(\Delta,J)$ takes the form that it does in (\ref{eq:rhoHatNaught}).\footnote{If you allow for $J<0$, then $\hat{\rho}_0(\Delta,J)$ generalizes to 
    \begin{align}
        \hat{\rho}_0(\Delta,J) &= [\delta_{J,0} - \delta_{J,-1}][\delta(\Delta)-\delta(\Delta-3)] - [\delta_{J,1} - \delta_{J,-2}][\delta(\Delta-1) - \delta(\Delta-2)].
    \end{align}
    This term is then invariant under the shadow transform, spin shadow transform, and light transform \cite{Kravchuk:2018htv}.
    } Note that for all subsequent analysis we will drop the identity contribution and focus on the prediction of non-trivial operators.

    With our closed-form expression in hand, it is first instructive to point out some interesting features. Returning to Fig. \ref{fig:rhoClosedForm}, we can now formally understand the disagreement between the Borel-Pad\'e resummed series and the closed-form answer. At any $J$,  the resummed answer and the closed-form expression disagree for $\Delta < J + 4$, which is exactly the threshold for all of the step functions in the closed-form expression (\ref{eq:DenOfStatesClosedForm}) to fully turn on. This observation is sensible since the perturbative large-$\Delta$ expansion is completely blind to the step functions. Additionally, the closed-form expression has unphysical negative behavior below the $\Delta = J + 4$ threshold for $J < 2$. For $J \geq 2$, this behavior goes away.\footnote{Strictly speaking, these are $f$-dependent statements, since (\ref{eq:DenOfStatesClosedForm}) depends on $f$. For the 3d Ising CFT, $J\geq 2$ is safe from this unphysical behavior. If $f < 0.135$, then $J=2$ picks up an unphysical region.}
    
    Let's now apply our closed-form density of primary states to the exact spectrum by seeing how well it predicts the count of CFT primary states by spin sector. Mechanically, we want to compare the observed count of primary states in a given spin-$J$ sector to the integrated density of states in that sector 
    \begin{align}
        N(\Delta,J) &= \int_0^\Delta d\Delta'\,\rho_\text{closed}(\Delta',J).
    \end{align}
    While this is a simple comparison, Fig. \ref{fig:rhoClosedForm} suggests that the closed-form expression may not be predictive below some threshold $\Delta$. We therefore would like to supply the count of states below some $\Delta_0$ as an input in order to see if it can improve agreement with data. Thus, we define
    \begin{align}
        N(\Delta,J;\Delta_0) &= N_\text{actual}(\Delta \leq \Delta_0,J) + \int_{\Delta_0}^{\Delta}d\Delta'\,\rho_\text{closed}(\Delta',J),\label{eq:AdjustedNumberOfStates}
    \end{align}
    with $N_\text{actual}(\Delta \leq \Delta_0, J)$ defined as
    \begin{align}
        N_\text{actual}(\Delta \leq \Delta_0, J) &= \sum_{\Delta_\mathcal{O}\leq\Delta_0}\theta(\Delta-\Delta_\mathcal{O})\delta_{J,J_\mathcal{O}}.
    \end{align}
    The idea is that we are using the exact spectrum up to $\Delta \leq \Delta_0$ and then the integrated closed-form density of states for $\Delta > \Delta_0$. Note that we will take $N_\text{actual}(\Delta,J)$ to mean the full count of states with no restriction $\Delta_0$.

\begin{figure}[h!]
        \centering
        \includegraphics[width=\textwidth]{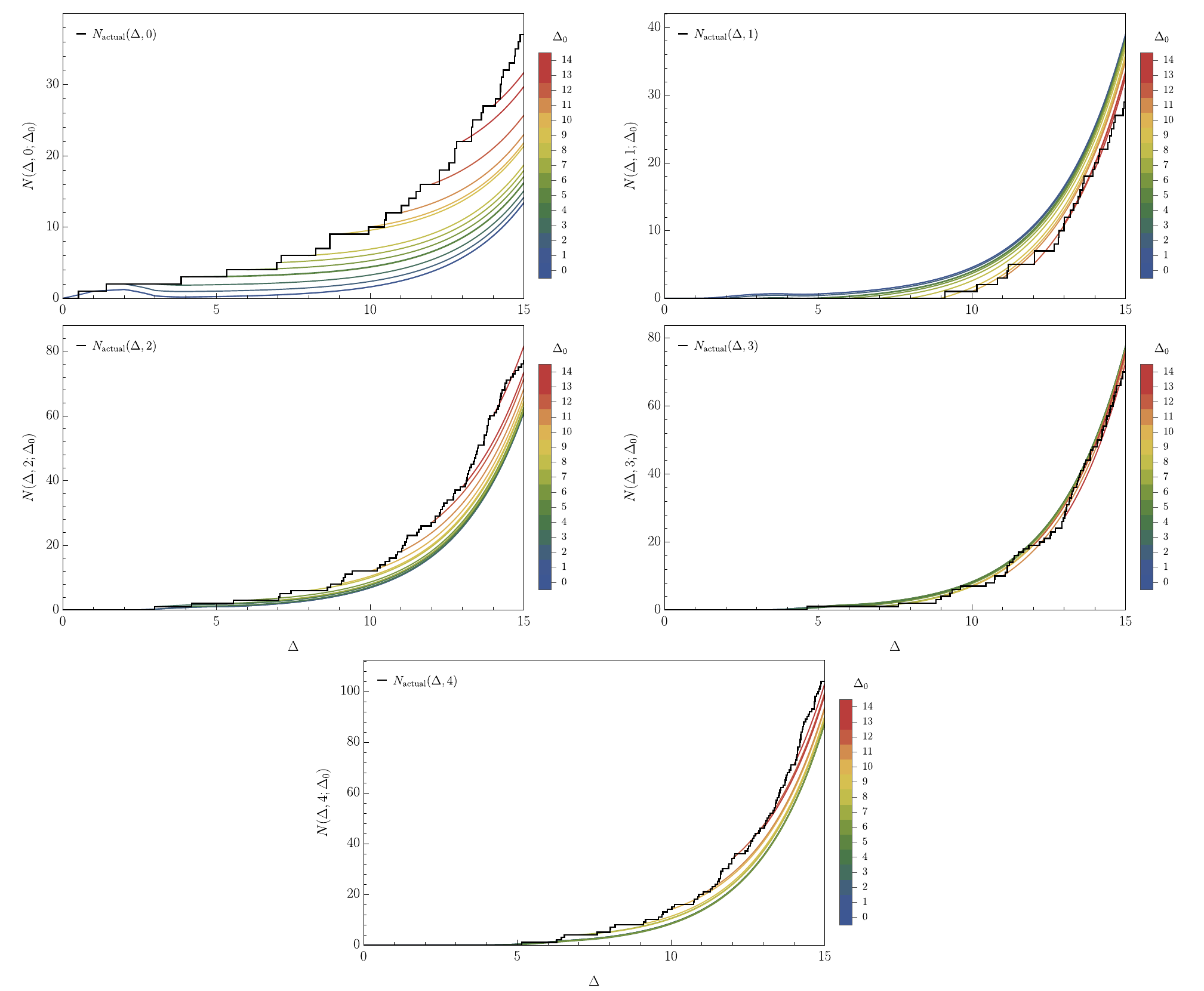}
        \caption{Predicted count of states $N(\Delta,J;\Delta_0)$ plotted against the actual count from the Fuzzy Sphere as a function of $\Delta$. Each panel is a fixed spin $J \leq 4$. Colors going from blue to red imply more non-perturbative input.}
        \label{fig:numOfOps}
    \end{figure}

    We now plot (\ref{eq:AdjustedNumberOfStates}) for each $J\leq4$ for all $\Delta < 15$ with a curve for each $\Delta_0 = 0,1,\ldots, 14$ in Fig. \ref{fig:numOfOps}. We observe that spin-0 is terribly under-predicted, spin-2 and spin-4 are mildly under-predicted, spin-1 does quite well once we supply the initial operators, and spin-3 agrees very well overall. These results are somewhat surprising, given that the spectrum still looks quite discrete even for $10 \leq \Delta < 15$. Perhaps even more surprising is how closely the predicted count tracks the actual count with minimal $\Delta_0$ input for some $J$.

    \subsection{Universal Subleading Corrections to ETH}\label{sec:UniversalSubleadingETH}

    Our goal now is to compute a tower of subleading corrections to the leading order ETH prediction, which is equivalently recast as a question about the asymptotic behavior of OPE coefficients. Our starting point is the thermal one-point function of a scalar primary operator $\phi$ on the sphere with non-zero chemical potential
    \begin{align}
        \langle \phi(x) \rangle_{\beta,\mu} &= \frac{1}{\mathcal{Z}(\beta,\mu)}\int_{0}^{\infty}d\Delta\sum_{J=0}^{\infty}\rho(\Delta,J)\sum_{a=0}^{J}\overline{\lambda^{a}_{\phi\mathcal{OO}}}(\Delta,J)g^{\Delta_\phi,a}_{\Delta,J}(\beta,\mu,\theta),
    \end{align}
    where we have taken $\Delta$ to be continuous and $\overline{\lambda^{a}_{\phi\mathcal{OO}}}(\Delta,J)$ is defined as an averaged OPE coefficient due to the possibility of degeneracies for fixed $(\Delta,J)$
    \begin{align}
        \overline{\lambda^{a}_{\phi\mathcal{OO}}} &= \frac{1}{n(\Delta,J)}\sum_{i=1}^{n(\Delta,J)}\lambda^a_{\phi\mathcal{O}_i\mathcal{O}_i}.
    \end{align}
    Applying the inversion formula developed in \cite{Buric:2025uqt} shows that the OPE coefficients are given by
    \begin{align}
        \overline{\lambda^{a}_{\phi\mathcal{OO}}}(\Delta,J) &= \frac{1}{\rho(\Delta,J)}\frac{1}{2\pi i}\int\mathcal{W}\,g^{3-\Delta_\phi,a}_{3-\Delta,J}(\beta,\mu,\theta)\mathcal{Z}(\beta,\mu)\langle\phi(x)\rangle_{\beta,\mu},\label{eq:OPECoeffDef}
    \end{align}
    where $\rho(\Delta,J)$ is the density of primary states as defined in Section \ref{sec:DensityOfPrimaryStates} and $\mathcal{W}$ an orthogonal measure on the space of thermal blocks 
    \begin{align}
        \int \mathcal{W} &= \int_{\gamma-i\infty}^{\gamma+i\infty}d\beta\int_{-\pi}^{\pi}d\mu\int_{0}^{\pi}d\theta\,\mathcal{W}(\beta,\mu,\theta),
    \end{align}
    with $\mathcal{W}(\beta,\mu,\theta)$ defined in terms of the orthogonal measure on the space of characters from (\ref{eq:MeasureForChi})
    \begin{align}
        \mathcal{W}(\beta,\mu,\theta) &= \omega(\beta,\mu)\frac{\sin\theta}{2}.
    \end{align}
    Since we are interested in the universal subleading behavior, we will again take the partition function to be the leading order term from the thermal effective action
    \begin{align}
        \mathcal{Z} &\approx e^{\frac{4\pi f}{\beta^2 + \mu^2}}.
    \end{align}
    For the same reason, we will take the sphere thermal one-point function to be the leading order answer one finds from the thermal effective action
    \begin{align}
        \langle \phi(x)\rangle_{\beta,\mu} \approx \frac{b_\phi}{(\beta^2 + \mu^2\sin^2\theta)^{\Delta_\phi/2}}.\label{eq:OnePtFuncBetaMu}
    \end{align}
    
    We now proceed in a similar fashion to Section \ref{sec:DensityOfPrimaryStates}. We will first evaluate (\ref{eq:OPECoeffDef}) by a saddle point approximation similar to \cite{Buric:2025uqt}, with the key difference being our use of (\ref{eq:OnePtFuncBetaMu}) as the form of the one-point function in the integrand. This procedure generates $\overline{\lambda^{a}_{\phi\mathcal{OO}}}$ as a series expansion in $\Delta^{-1/3}$. We then go further by resumming the series using Borel-Pad\'e resummation. Before proceeding, note that (\ref{eq:OPECoeffDef}) is a general prediction about all OPE coefficients for any tensor structure. However, our interest in ETH makes us narrowly interested in the $a=0$ case. Thus, we restrict to the $a=0$ tensor structure and define
    \begin{align}
        \overline{f_{\phi\mathcal{OO}}} &\equiv \overline{\lambda^{a=0}_{\phi\mathcal{OO}}}.
    \end{align}
    Further, we drop the $a$ index from all thermal blocks and it should be understood that we are considering the $a=0$ tensor structure unless stated otherwise.

    In order to carry out the saddle point analysis, we must deal with the shadow thermal blocks appearing in the integrand of (\ref{eq:OPECoeffDef}). Rather than approximating the blocks as a low-temperature expansion, \cite{Buric:2025uqt} showed that by defining
    \begin{align}
        g^{\Delta_\phi}_{\Delta,J}(q,y,s) &= q^{\Delta}F^{\Delta_\phi}_{\Delta,J}(q,y,s),
    \end{align}
    $F^{\Delta_\phi}_{\Delta,J}(q,y,s)$ can be computed recursively as a large-$\Delta$ series expansion where each term is all orders in $q$. Although this works for any spin $J$, it will be too computationally expensive to carry out for generic $J$. Thus, we compute the necessary series on a case-by-case basis for $J \leq 4$. Further details on the computation of these large-$\Delta$ thermal blocks can be found in \cite{Buric:2025uqt}.

    Putting everything together, the integral we want to compute via saddle point is of the form
    \begin{align}
        \overline{f_{\phi\mathcal{OO}}} &= \frac{b_\phi}{\rho(\Delta,J)}\frac{1}{2\pi i}\int_{\gamma-i\infty}^{\gamma+i\infty}d\beta\int_{-\pi}^{\pi}d\mu\int_0^\pi d\theta H(\beta,\mu,\theta)e^{\frac{4\pi f}{\beta^2+\mu^2}+\beta(\Delta-3)},
    \end{align}
    where we have defined 
    \begin{align}
        H(\beta,\mu,\theta) &\equiv \frac{\mathcal{W}(\beta,\mu,\theta)}{(\beta^2+\mu^2\sin^2\theta)^{\Delta_\phi/2}}\sum_{m=0}^{\infty}(3-\Delta)^{-m}\left(F^{3-\Delta_\phi}_{3-\Delta,J}\right)^{(m)}(\beta,\mu,\theta).
    \end{align}
    From here, we follow along the same procedure outlined in Section \ref{sec:DensityOfPrimaryStates}, again working with the variables $(x_L,x_R)$ that were defined in (\ref{eq:xLxR}). The only added final step is dividing by the density of states series expansion (\ref{eq:DenOfStatesSeries}). 

    In the end, we carried out the saddle point analysis through $\mathcal{O}(\Delta^{-19/3})$ relative to the leading order term for all $J \leq 4$, which is equivalent to saying that we generated 20 terms of the series.\footnote{This counting of terms includes the missing $\mathcal{O}(\Delta^{-1/3})$ which happens to have an identically zero series coefficient.} Below are the first few terms in the expansion
    \begin{align}
        \overline{f_{\phi\mathcal{OO}}}(\Delta,J) &= b_\phi\left(\frac{\Delta}{8\pi f}\right)^{\Delta_\phi/3}\Bigg[1 + \frac{\Delta_\phi(6- \Delta_\phi)}{3\pi^{1/3}f^{1/3}\Delta^{2/3}} + \frac{\Delta_\phi(\Delta_\phi-5)}{4\Delta}\label{eq:ETHCorrectionseries} \\
        &+ \frac{\Delta_\phi(108 - 9\pi f(\Delta_\phi - 5) + 174\Delta_\phi - 68\Delta_\phi^2 + 6\Delta_\phi^3)}{108\pi^{2/3}f^{2/3}\Delta^{4/3}} + \mathcal{O}(\Delta^{-5/3})\Bigg].\nonumber
    \end{align}
    Note that higher order terms have been suppressed for the sake of space, and the first $J$-dependent contribution does not turn on until $\mathcal{O}(\Delta^{-2})$ and takes the form $-\frac{\Delta_\phi J(J+1)}{3\Delta^2}$.
    
\begin{figure}[h!]
        \centering
        \includegraphics[width=\linewidth]{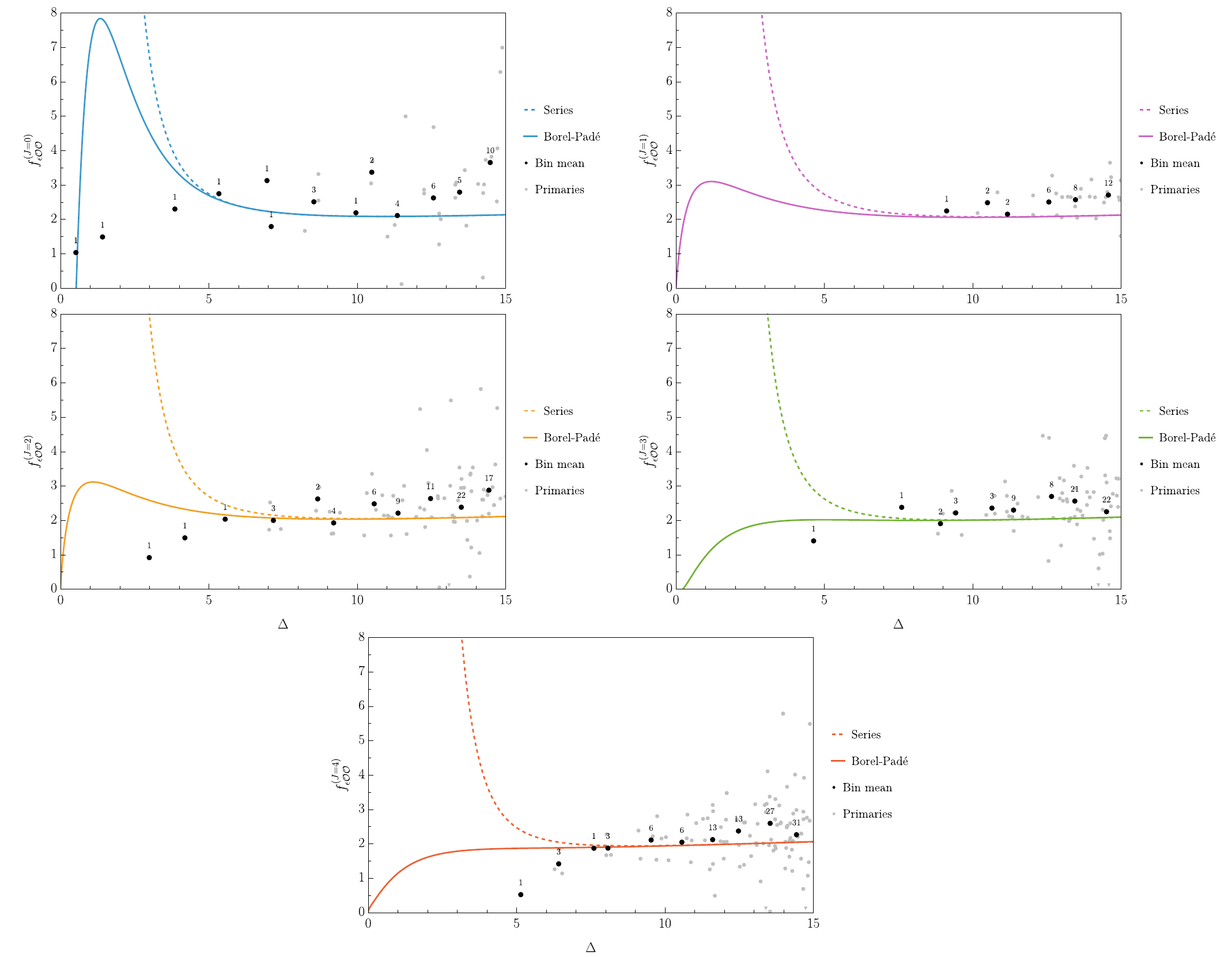}
        \caption{
        ETH plus 20 subleading terms and its Borel-Pad\'e resummation plotted as a function of $\Delta$. Each panel is a fixed spin $J\leq 4$. Raw OPE data is plotted in gray, with unit width bin means of the data plotted in black. Arrows indicate a point outside of frame. Horizontal positions of the bin means are set by the mean dimension of its constituents.
        }
        \label{fig:ETHAllSubleadingTermsAndBorel}
    \end{figure}

    We now repeat the exercise from Section \ref{sec:FSandETH} and compare ETH with subleading universal corrections to the OPE data from the 3d Ising CFT. We plot the data and its corresponding series expansion in Fig. \ref{fig:ETHAllSubleadingTermsAndBorel}. For this comparison we average OPE data in bins of unit width, where we align the bin edges with integers. Each bin mean is placed at the mean dimension of its constituents.
    
    We observe that beyond $\Delta \gtrsim 9$, the OPE data is consistent with the series expansion (\ref{eq:ETHCorrectionseries}). This agreement improves with increasing spin, possibly due to the number of states growing with spin and giving us better statistics. The $J=0$ data seems to do the worst, but this is not inconsistent with the count of CFT states comparisons from Section \ref{sec:RhoExactBonus}, where the agreement with scalar primaries was also the worst.

    As it stands, the series expansion (\ref{eq:ETHCorrectionseries}) does quite well at predicting the microcanonically averaged OPE data beyond $\Delta \gtrsim 9$. However, since (\ref{eq:ETHCorrectionseries}) is still in a pre-asymptotic regime, it is interesting to ask whether we can do better and describe data at even lower dimensions by adding more terms. A related concern is the fact that regardless of the number of terms, the series unphysically diverges at small $\Delta$. Motivated by our observations with the density of primary states in Section \ref{sec:RhoExactBonus}, we turn to Borel-Pad\'e resummation to address both issues.
    
    The Borel-Pad\'e resummed versions of each ETH series are plotted in Fig. \ref{fig:ETHAllSubleadingTermsAndBorel}. We observe that the resummed curves all agree with their corresponding parent series expansion for $\Delta \gtrsim 9$, exactly where we observed good agreement between the series and the averaged OPE data. Slightly below this threshold the resummed curves and series split apart for $J \geq 1$, staying finite and modestly sized at small $\Delta$. The $J=0$ curve mirrors this behavior, but does not split apart until $\Delta\approx 5$. Importantly, the Borel-Pad\'e resummed curves do not significantly improve the range of $\Delta$ over which each ETH series is predictive. While this is disappointing, it is not at all inconsistent. Further, the data in these low-$\Delta$ regimes is easily accessible by non-perturbative methods, so there is no practical need to predict this data thermally. The important takeaway is that microcanonical averages of these OPE coefficients of operators with $\Delta\in[10,15)$ are already well-described by a thermal ensemble.
    
\section{Summary and Future Directions}\label{sec:SummaryAndFutureDirec}

    In this work we have explored the relationship between CFT data and thermal observables in 3d CFTs. On one hand, we investigated how well we can predict CFT data from thermal observables using ETH and universal subleading corrections. On the other hand, we investigated how well we could approximate thermal observables from the low-dimension CFT data.

    Our key finding in the forward direction is that microcanonical averages of diagonal OPE coefficients are well-described by a thermal distribution starting as early as $\Delta \gtrsim 9$. Our key result in the converse direction is a new estimate of the free energy parameter $f$ and thermal one-point function $b_\epsilon$ solely using the collection of CFT data computed from the Fuzzy Sphere, which agree with previous results \cite{PhysRevE.53.4414,PhysRevE.56.1642,PhysRevE.79.041142,Iliesiu:2018zlz,Barrat:2025wbi}. Along the way, we computed a closed-form expression for the density of primary states, which in turn allowed us to predict the count of primary states in the 3d Ising CFT to unexpectedly decent accuracy. All direct data comparisons were made possible by having computed CFT data in the Fuzzy Sphere for 319 primary operators in the 3d Ising CFT, most of which were not previously known.

    From here, there are many interesting directions to explore. One obvious path would be attempting to apply the type of analysis we have considered here to other 3d CFTs. Alternatively, it would be interesting to consider $\epsilon'$, the first irrelevant $\mathbb{Z}_2$-even primary operator, and apply our analysis. One could attempt to extract its corresponding thermal vev $b_{\epsilon'}$, and then subsequently see how well its diagonal OPE coefficients are described by a thermal distribution in a similar range of $\Delta$.

    Another interesting extension would be an exploration of other thermal quantities such as two-point functions. The sphere thermal two-point function can again be understood as a Boltzmann sum over CFT data, but now with off-diagonal OPE coefficients. A simple example in the 3d Ising CFT would be the thermal two-point function of $\sigma$ on the sphere, $\langle \sigma(x)\sigma(0)\rangle_{S^1_\beta\times S^2}$. Carrying out the calculation in this way would allow one to investigate the dynamics outside of the regime where the OPE converges.

    Yet another interesting direction would be exploring thermal observables directly in a Fuzzy Sphere setup. Taking the 3d Ising CFT as an example, it would be interesting to use the operators and eigenstates from the Fuzzy Sphere to compute the thermal one-point function of $\epsilon$ on the sphere directly in order to compare with what we found here. If this method turned out to be more accurate, it would be much simpler than exporting out all of the data and performing the analysis as we did.

    Finally, there are two further mysteries worth exploring which are not specifically related to thermal physics. The full set of diagonal OPE coefficients of $\epsilon$ in Fig. \ref{fig:LeadingNaiveETH} displays a number of interesting features which are unexplained. For example, when we organize the OPE coefficients by spin there seem to be some spin-dependent trajectories, which are particularly apparent when $\Delta$ is small and spin is $J=0$ or 2. It is possible that this structure can be understood by identifying this data with the semiclassical $\phi^n$ type states identified in \cite{Zhou:2026ugg}, but it would be interesting to make this concrete. Another puzzling feature of the data is the appearance of what looks like three different `bands' that some of the OPE data clusters along. It turns out that these bands are organized by $\mathbb{Z}_2$ charge of the external states. It would be interesting to understand this tantalizing feature of the data and provide some explanation for this organization.
    
\acknowledgments
We thank Giulia Fardelli, Ami Katz, Alessandro Vichi, and Zheng Zhou for helpful discussions. Some numeric results in this paper used the publicly available Julia code provided at \href{https://www.fuzzified.world}{https://www.fuzzified.world} and were performed on the Shared Computing Cluster which is administered by Boston University’s Research Computing Services.  Claude (Opus 5.5 and Fable 5.1) was used for writing code, running code, verifying formulae, and putting together the ancillary file. ALF and NR are supported by the US Department of Energy Office of Science under Award Number DE-SC0015845.

\appendix

\section{An ``Exact'' Density of Primary States}\label{app:rhoExact}

    As was mentioned in Section \ref{sec:RhoExactBonus}, we can directly evaluate (\ref{eq:DensityOfStatesDef}) without any approximations. Our starting point is again (\ref{eq:rhoIntExpExpanded}) which we repeat below for convenience
    \begin{align}
        \rho(\Delta,J) &= \sum_{n=0}^{\infty}\frac{(4\pi f)^n}{n!}\frac{1}{2\pi i}\int_{\gamma-i\infty}^{\gamma+i\infty}d\beta \int_{-\pi}^{\pi}d\mu\,\omega(\beta,\mu)\chi_{3-\Delta,J}(\beta,\mu)\frac{1}{(\beta^2+\mu^2)^n}.
    \end{align}
    We also simplify the product of the measure and character down to the same simple sums as in (\ref{eq:simplifiedRhoIntegrand}), which we again repeat below for convenience
    \begin{align}
        \frac{1}{4\pi}\sum_{s=\pm 1}\left[\sum_{\delta\in\{0,3\}}\sum_{j\in\{0,1\}} - \sum_{\delta\in\{1,2\}}\sum_{j\in\{-1,2\}}\right](-1)^{\delta+j}e^{\beta(\Delta-\delta)}e^{i\mu s(J+j)}.
    \end{align}
    Therefore, we will focus on evaluating 
    \begin{align}
        \frac{1}{2\pi i}\int_{\gamma-i\infty}^{\gamma+i\infty}d\beta\int_{-\pi}^{\pi} d\mu\,\frac{e^{\beta \overline{\Delta}}e^{i\mu\overline{J}}}{(\beta^2+\mu^2)^n},
    \end{align}
    where $\overline{\Delta} = \Delta - \delta$ and $\overline{J} = s(J + j)$. Finally, recall that in Section \ref{sec:RhoExactBonus} we already evaluated the $n=0$ term exactly, so from here on we focus on $n > 0$.

    Our first new step is to swap the order of integration and compute the $\beta$-integral first, yielding 
    \begin{align}
        \frac{1}{2\pi i}\int_{\gamma-i\infty}^{\gamma+i\infty}d\beta\, \frac{e^{\beta\overline{\Delta}}e^{i\mu\overline{J}}}{(\beta^2+\mu^2)^n} &= \frac{\sqrt{\pi}e^{i\mu\overline{J}}\overline{\Delta}^{n-\frac{1}{2}}}{2^{n-\frac{1}{2}}\Gamma(n)}\frac{J_{n-\frac{1}{2}}(\overline{\Delta}\mu)}{\mu^{n-\frac{1}{2}}}\theta(\overline{\Delta}),
    \end{align}
    where $J_{\alpha}(x)$ is a Bessel function of the first kind. It is then convenient to rewrite $J_{\alpha}(x)$ as a spherical Bessel function of the first kind $j_{\alpha}(x)$
    \begin{align}
        J_{n-\frac{1}{2}}(\overline{\Delta}\mu) &= \sqrt{\frac{2\overline{\Delta} \mu}{\pi}}j_{n-1}(\overline{\Delta} \mu),
    \end{align}
    since $j_\alpha(x)$ has the following nice integral representation 
    \begin{align}
        j_\alpha(x) &= \frac{x^\alpha}{2^{\alpha+1}\alpha!}\int_{-1}^{1}dt\,(1-t^2)^{\alpha} e^{ixt}.
    \end{align}
    Thus, the $\beta$-integral has left us with 
    \begin{align}
        \frac{1}{2\pi i}\int_{\gamma-i\infty}^{\gamma+i\infty}d\beta\, \frac{e^{\beta\overline{\Delta}}e^{i\mu\overline{J}}}{(\beta^2+\mu^2)^n} &= \left[\frac{\overline{\Delta}^{\,n-\frac{1}{2}}}{2^{n-\frac{1}{2}}\Gamma(n)}\right]^2\int_{-1}^{1}dt\,(1-t^2)^{n-1}e^{i\mu(\overline{\Delta} t + \overline{J})}\theta(\overline{\Delta}).
    \end{align}
    The $\mu$-integral is now trivial if we swap it with the $t$-integral, and we have shown that the integral we wanted to evaluate is given by 
    \begin{align}
        \frac{1}{2\pi i}\int_{\gamma-i\infty}^{\gamma+i\infty}d\beta\int_{-\pi}^{\pi} d\mu\,\frac{e^{\beta \overline{\Delta}}e^{i\mu\overline{J}}}{(\beta^2+\mu^2)^n} &= \left[\frac{\sqrt{\pi}\,\overline{\Delta}^{\,n-\frac{1}{2}}}{2^{n-1}\Gamma(n)}\right]^2\int_{-1}^{1}dt\,(1-t^2)^{n-1}\text{sinc}(\pi(\overline{\Delta}t + \overline{J}))\theta(\overline{\Delta}).
    \end{align}
    The final step to push this calculation as far as possible analytically is to now swap the summation over $n$ with the $t$-integral and resum all of the terms 
    \begin{align}
        \sum_{n=1}^{\infty}\frac{(4\pi f)^n}{n!}\left[\frac{\sqrt{\pi}\,\overline{\Delta}^{\,n-\frac{1}{2}}}{2^{n-1}\Gamma(n)}\right]^2(1-t^2)^{n-1} &= 4\pi^2 f \Delta {}_0F_2\left(;1,2;\pi f(1-t^2)\Delta^2\right).
    \end{align}

    Collecting everything together, we have therefore shown that the ``exact'' density of primary states is given by 
    \begin{align}
        \rho_\text{exact}(\Delta,J) &= \hat{\rho}_0(\Delta,J) + \left[\sum_{\delta\in\{0,3\}}\sum_{j\in\{0,1\}} - \sum_{\delta\in\{1,2\}}\sum_{j\in\{-1,2\}}\right](-1)^{\delta+j}\hat{\rho}_\text{exact}(\Delta-\delta,J+j),
    \end{align}
    where $\hat{\rho}_0(\Delta,J)$ is given by (\ref{eq:rhoHatNaught}) and $\hat{\rho}_\text{exact}(\Delta,J)$ is defined as 
    \begin{align}
        \hat{\rho}_\text{exact}(\Delta,J) &\equiv 2\pi f \Delta\int_{-1}^{1}dt\, {}_0F_2\left(;1,2;\pi f \Delta^2(1-t^2)\right)\text{sinc}(\pi(\Delta t + J))\theta(\Delta).\label{eq:exactRhoSeed}
    \end{align}
    
    A nice consistency check is showing that we recover the closed-form expression (\ref{eq:RhoClosedFormSeed}) by taking the appropriate $\Delta\to\infty$ and $J=\text{fixed}$ limit of (\ref{eq:exactRhoSeed}). To see this, we first notice that in the $\Delta\to\infty$ and $J=\text{fixed}$ limit we have
    \begin{align}
        \Delta\text{sinc}(\pi(\Delta t+J)) \xrightarrow[]{\quad} \delta(t-t_0)\quad \text{with} \quad t_0 \equiv -\frac{J}{\Delta}.
    \end{align}
    We then observe that we can rewrite the integral (\ref{eq:exactRhoSeed}) as
    \begin{align}
         2\pi f \Delta \int_{-\infty}^{\infty} dt\,\theta(1-|t|){}_0F_2\left(;1,2;\pi f\Delta^2(1-t^2)\right)\text{sinc}(\pi(\Delta t+J))\theta(\Delta).
    \end{align}
    Taking the limit of the above rewritten integral leaves us with 
    \begin{align}
        2\pi f {}_0F_2\left(;1,2;\pi f\Delta^2(1-t_0^2)\right) \theta(1-|t_0|) \theta(\Delta) &= 2\pi f {}_0F_2\left(;1,2;\pi f(\Delta^2-J^2)\right)\theta(\Delta-|J|),
    \end{align}
    which is precisely (\ref{eq:RhoClosedFormSeed}).

    \begin{figure}[h!]
        \centering
        \includegraphics[width=0.9\linewidth]{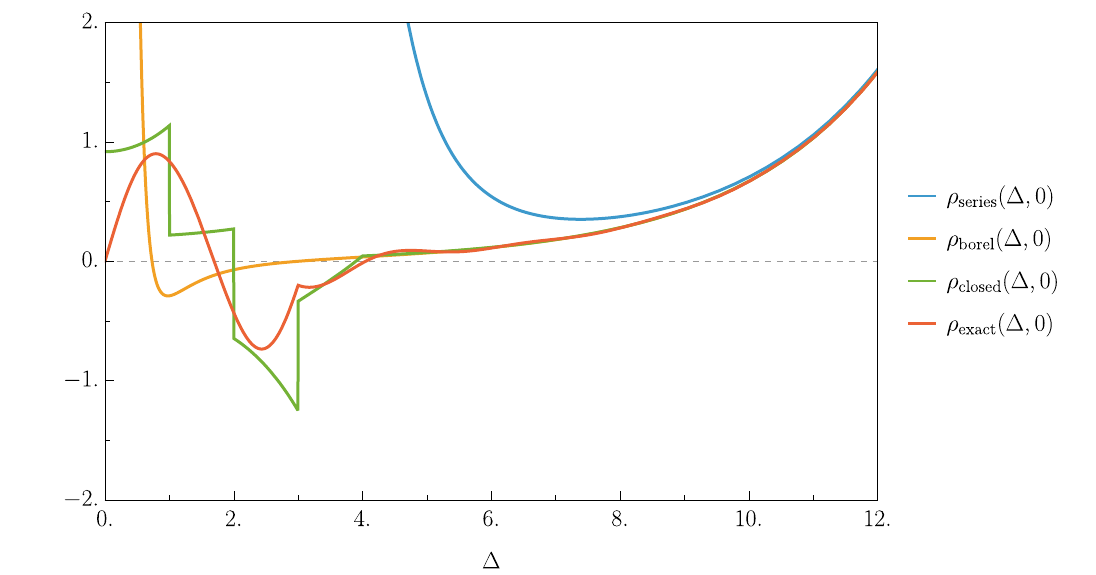}
        \caption{Comparison of the spin-0 density of primary states as: a 22 term asymptotic series, a Borel-Pad\'e resummation, a closed-form expression, and an ``exact'' expression.}
        \label{fig:rhoExact}
    \end{figure}

    We can now make a final comparison between the perturbative series, Borel-Pad\'e resummation, the closed-form expression, and now the ``exact'' expression. We plot the collection of these curves in Fig. \ref{fig:rhoExact}, again for $J=0$. We observe that down to $\Delta \approx 4$, all attempts to resum or complete the series agree quite well. The main difference with the ``exact'' expression is mild oscillation layered on top of the curve above $\Delta \approx 4$ which dies away by $\Delta \approx 8$. Just like the closed-form expression, the ``exact'' one becomes unphysical below $\Delta \approx 4$, with the behavior in this region behaving like a smoothed out version of the closed-form expression. These observations are repeated for $J > 0$.

    There are a couple of key takeaways from this comparison. First, there seems to be no further information that can be extracted from this calculation below $\Delta\approx 4$. As with all of the previous calculations, there is no contradiction given that the partition function we chose in the integrand is a high-temperature approximation and we are pushing into a regime where it should in principle be invalid. However, this leads to the second key takeaway that the high-temperature partition function seems to be a good approximation down to quite small $\Delta$.

\section{Borel-Pad\'e Resummation}\label{app:BorelPade}

    In this section we enclose the details of the Borel-Pad\'e resummations for the density of primary states and ETH, which were used in Sections \ref{sec:DensityOfPrimaryStates} and \ref{sec:UniversalSubleadingETH} respectively. Although the series expansions we derived via saddle point approximation have an insufficient number of terms to show their asymptotic nature, the closed-form density of primary states (\ref{eq:DenOfStatesClosedForm}) can be expanded to establish that the series is asymptotic with the following growth\footnote{To be able to observe this asymptotic growth, we find that we must have $n\gtrsim 50$ for spin-0 with the minimum $n$ growing with $J$.}
    \begin{align}
        \left\vert \frac{c_{n+1}}{c_{n}}\right\vert &\sim \sqrt{n}, & 
        \left\vert \frac{c_{n+2}}{c_{n}}\right\vert &\sim n.\label{eq:coeffScalingBorelPade}
    \end{align}
    This growth suggests that the natural variable for understanding this series is $\Delta^{-2/3}$. Physically this makes sense since the saddle is set by the thermodynamic entropy which goes as $S\propto \Delta^{2/3}$, and the series generates fluctuation terms which are suppressed by powers of the entropy.\footnote{One can go even further and observe that terms in the series are not only suppressed by powers of the entropy evaluated at the saddle, but also powers of the temperature evaluated at the saddle noting that $T\propto \Delta^{1/3}$. Thus, these series expansions can in principle be organized by powers of $S$ and $T$.} Since the saddle structure is set by the leading term in the thermal effective action for both the density of states and ETH, we can conclude that the ETH series is likely asymptotic with the same large-$n$ coefficient scaling (\ref{eq:coeffScalingBorelPade}). Given this asymptotic scaling, we use $\Gamma(\frac{n}{2} + 1)$ for the Borel transform.
    
    For the density of primary states series (\ref{eq:DenOfStatesSeries}), we define 
    \begin{align}
        \hat{\rho}(\Delta,J)\equiv \frac{\rho(\Delta,J)}{\frac{8\pi^{2/3}f^{5/3}(2J+1)}{\sqrt{3}\Delta^{11/3}}e^{3\pi^{1/3}f^{1/3}\Delta^{2/3}}} = \sum_{n=0}^{21}c_nx^n,
    \end{align}
    where $x\equiv \Delta^{-1/3}$ and for clarity we have $c_0 = 1,c_1 = - 3\pi^{1/3}f^{1/3},$ and so on. We then Borel transform the series 
    \begin{align}
        \mathcal{B}(t) &= \sum_{n=0}^{21}\frac{c_n}{\Gamma(\frac{n}{2}+1)}t^n,
    \end{align}
    followed by analytically continuing the Borel transformed series using a Pad\'e approximant 
    \begin{align}
        \mathcal{PB}(t) &\equiv \mathcal{B}^{[10/11]}(t).
    \end{align}
    We proceed by computing the so-called lateral Borel resummation in order to avoid any singularities on the real axis
    \begin{align}
        B_{\theta_\pm}(x) &= \int_{\mathcal{C}_{\theta_\pm}}e^{-t}\mathcal{PB}(\sqrt{t}x) = e^{\pm i\theta}\int_{0}^{\infty}dt\,e^{-te^{\pm i\theta}}\mathcal{PB}(\sqrt{t}e^{\pm i\theta/2}x),
    \end{align}
    where $\mathcal{C}_{\theta_\pm}$ is a contour rotated above/below the real-$t$ axis by an angle $\theta$. We can make this a little nicer to work with by defining $t = u^2$ and $\varphi = \theta/2$ giving us 
    \begin{align}
        B_{\varphi_\pm}(x) &= 2e^{\pm 2i\varphi}\int_0^\infty du\,ue^{-u^2e^{\pm 2i\varphi}}\mathcal{PB}(ue^{\pm i\varphi}x).
    \end{align}
    The final Borel resummed result is then the average of the above integral along the two reflected contours
    \begin{align}
        B(x) &= \frac{1}{2}(B_{\varphi_+}(x) + B_{\varphi_-}(x)) = \text{Re}[B_{\varphi_\pm}(x)],\label{eq:FinalBorelRealPart}
    \end{align}
    which is equivalent to taking the real part of just one of the two integrals.

    We can follow an identical procedure for the ETH series (\ref{eq:ETHCorrectionseries}) with $J \leq 4$. We first define 
    \begin{align}
        \hat{f}_{\epsilon\mathcal{OO}}(\Delta,J) &\equiv \frac{\overline{f_{\epsilon\mathcal{OO}}}(\Delta,J)}{b_\epsilon\left(\frac{\Delta}{8\pi f}\right)^{\Delta_\epsilon/3}} = \sum_{n=0}^{19}c_nx^n,
    \end{align}
    where $x \equiv \Delta^{-1/3}$ and we note for clarity that $c_0 = 1$, $c_1 = 0$, $c_2 = \frac{\Delta_\epsilon(6-\Delta_\epsilon)}{3\pi^{1/3}f^{1/3}}$, and so on. As before we Borel transform the series 
    \begin{align}
        \mathcal{B}_J(t) &= \sum_{n=0}^{19}\frac{c_n}{\Gamma(\frac{n}{2}+1)}t^n,
    \end{align}
    followed by generating its Pad\'e approximant 
    \begin{align}
        \mathcal{PB}_J(t) &\equiv \mathcal{B}_J^{[9/10]}(t),
    \end{align}
    where we point out that the same Pad\'e order is chosen for all $J$. We then perform the same lateral Borel resummation procedure in order to generate the curves $B_J(x)$.

\section{Fuzzy Sphere Data Details}\label{app:FuzzySphereData}

   Here we provide brief explanations on choices we made when collecting and extrapolating the CFT data, omitting technical details of the methods themselves since they are already well described in \cite{Fardelli:2026zas} and \cite{Zhou:2026ugg}.

    For our computations we used the FuzzifiED Julia code and its built-in $SO(3)$-resolved basis. At each spin $J \leq 4$ and within each symmetry sector, we computed energy eigenstates up to finite-$N$ energies of $\Delta_\text{raw}\approx 17$ in order to decrease the likelihood of truncation effects distorting operators near the top of the spectrum. We performed this calculation for $12 \leq N \leq 18$ (including odd $N$). Primaries were identified using $|K|^2$, with our construction of $K_\mu$ matching that of \cite{Fardelli:2026zas} rather than that of \cite{Zhou:2026ugg}.\footnote{In \cite{Fardelli:2026zas} matrix elements of $K_\mu$ are identified in $\Lambda_\mu$ via its action on states, whereas \cite{Zhou:2026ugg} uses the conformal algebra to pick off $K_\mu$ from $\Lambda_\mu$. The former is more computationally expensive but yields much smaller $|K|^2$ eigenvalues, while the latter is computationally cheaper but yields larger $|K|^2$ eigenvalues.} Diagonal OPE coefficients of $\epsilon$ were computed according to \cite{Fardelli:2026zas}, where we also follow their prescription for subtracting off the contribution of $T^{00}$.

    For the large-$N$ extrapolations of the CFT data, we first computed a linear $1/N$ fit over the four best points $N=15$-18, yielding an extrapolated value $\Delta_{N=15\text{-}18}$. In order to estimate error bars, we then computed another linear $1/N$ fit which instead included all available data points, yielding an extrapolated value $\Delta_{N\leq 18}$. The data is reported as
    \begin{align}
        \Delta_{N=15\text{-}18} \pm | \Delta_{N=15\text{-}18} - \Delta_{N\leq 18} |.
    \end{align}
    This same procedure was carried out for the OPE coefficients as well. Note that three spin-4 operators at the top of the spectrum have no error attached since they only have data for $N\geq 15$.
    
    In the end, we were able to extract a total of 319 primaries with $\Delta < 15$ across spin sectors $J \leq 4$. Importantly, 92 of these primaries were known to the Fuzzy Sphere prior to this work, and primaries with $\Delta \geq 15$ which were recently identified in \cite{Zhou:2026ugg} are not included in our set. All of the raw and extrapolated data we computed is collected in an ancillary file for download.

    Finally, we briefly discuss a technical point in the microcanonical averaging of OPE data in Section \ref{sec:UniversalSubleadingETH}. A handful of states (six spin-2, two spin-3, ten spin-4) experience level repulsion at finite-$N$, leading to poor OPE coefficient extrapolations. We therefore average such pairs at finite-$N$ and extrapolate these averages with the previously described procedure. Each operator in a mixing pair is assigned this averaged OPE coefficient.

\nocite{*}
\bibliographystyle{JHEP}
\bibliography{bibliography.bib}

\end{document}